\documentclass[aps,prd, 12pt, notitlepage,a4paper,floatfix,showpacs,nofootinbib,superscriptaddress,tikz]{revtex4-2}

\pdfoutput=1
\usepackage[usenames,dvipsnames]{color}  
\usepackage{graphicx} 
\RequirePackage[colorlinks=true,urlcolor=red,anchorcolor=red,citecolor=red,filecolor=red,linkcolor=red,menucolor=red,pagecolor=red,linktocpage=true,pdfproducer=medialab,pdfa=true]{hyperref}

\usepackage[compat=1.1.0]{tikz-feynman}
\usepackage{comment}
\usepackage{xcolor}
\usepackage{cancel}
\usepackage{bm}
\usepackage{bbold}
\usepackage{graphicx}
\usepackage{latexsym}
\usepackage{amsthm}
\usepackage{amsmath}
\usepackage{tabularx}
\usepackage{placeins}
\usepackage{wrapfig}
\usepackage{amssymb}
\usepackage{cancel}
\usepackage{multirow}
\usepackage{pst-3d}                  
\usepackage{pst-grad}                
\usepackage{pst-node}                
\usepackage{pst-slpe}                
\usepackage{pst-plot}                %
\usepackage{pst-coil}  %
\usepackage{pst-math}                %
\usepackage{pstricks-add}            %
\usepackage{rotating}
\usepackage{geometry}
\usepackage{slashed}
\usepackage{hhline}
\usepackage{float}
\usepackage{fancybox}
\usepackage[utf8]{inputenc}
\usepackage[normalem]{ulem} 
\usepackage[normalem]{ulem}

\usepackage{color,soul}
\setul{0.5ex}{0.3ex}
\setulcolor{blue}

\usepackage{braket}

\definecolor{linkcolor}{rgb}{0,0,0.5}
\allowdisplaybreaks
\allowdisplaybreaks[1]
\allowdisplaybreaks[2]

\definecolor{greenLinks}{rgb}{0, 0.6, 0}
\definecolor{blueLinks}{rgb}{0, 0, 0.6}
\definecolor{redLinks}{rgb}{0.6, 0, 0}
\definecolor{tempText}{rgb}{0.55, 0.10,0.67}
\definecolor{eprintLinks}{rgb}{0.4, 0.4, 0.4}
\definecolor{journalLinks}{rgb}{0.6, 0, 0}
\newcommand {\ignore}[1]{}

\usepackage{soul}
\definecolor{mightnightblue}{RGB}{25,25,112}

\definecolor{brown}{rgb}{0.59, 0.29, 0.0}

\definecolor{darkred}{rgb}{0.6,0,0}

\def\lsim{\mathrel{\rlap{\lower4pt\hbox{\hskip1pt$\sim$}}
    \raise1pt\hbox{$<$}}}
\def\gsim{\mathrel{\rlap{\lower4pt\hbox{\hskip1pt$\sim$}}
    \raise1pt\hbox{$>$}}}

\newcommand{\AddrBhopal}{Department of Physics, Indian Institute of Science Education and Research - Bhopal, Bhopal Bypass Road, Bhauri, Bhopal 462066, INDIA}

\newcommand{\AddrDharamshala}{Department of Physics and Astronomical Science, Central University of Himachal Pradesh, Dharamshala 176215, INDIA}

\begin{document}

\title{Type-III Scotogenic Model: Inflation, Dark Matter\\ and Collider Phenomenology}
\author{Labh Singh}\email{sainilabh5@gmail.com}
\affiliation{\AddrDharamshala}
\author{Rahul Srivastava}\email{rahul@iiserb.ac.in}
\affiliation{\AddrBhopal}
\author{Surender Verma }\email{s\_7verma@hpcu.ac.in}
\affiliation{\AddrDharamshala}
\author{Sushant Yadav}\email{sushant20@iiserb.ac.in}
\affiliation{\AddrBhopal}


\begin{abstract}
\vspace{2cm}

\noindent We explore an extension of the Type-III scotogenic model through the inclusion of a real scalar singlet, which play the role of the inflaton by virtue of its non-minimal coupling to the Ricci scalar. This scalar field not only drives inflation but also decays into other particles embedded within the Type-III scotogenic framework. Within this theoretical structure, both the inert scalar doublet and the fermionic triplet are instrumental in radiatively generating neutrino masses, while, also, emerging as viable candidates for dark matter component of the Universe. We conduct a detailed analysis of their relic abundance and assess their prospects for direct detection. Additionally, we examine the potential collider signatures associated with this framework, emphasizing avenues for experimental verification in forthcoming high energy physics experiments. Owing to the suppressed production cross-section of the fermionic triplet at the LHC, we extend our analysis to the Future Circular Collider in its hadron-hadron mode (FCC-hh). Within this context, we identify the mono-leptonic decay channel of the triplet fermion, accompanied by substantial missing transverse energy, as a salient and experimentally distinguishable signature of the model. Furthermore, our study reveals that the leading-order production cross-section of low-mass scalar dark matter lies within the prospective sensitivity reach of both leptonic and hadronic collider experiments, thereby offering promising avenues for experimental scrutiny.

\end{abstract}
\maketitle
\newpage
\tableofcontents
\newpage
\section{Introduction}
\noindent The experiments in neutrino oscillation physics \cite{ParticleDataGroup:2024cfk} and modern cosmology \cite{Planck:2018vyg} have, among others, led to the establishment of two important facts about the working of nature (i) that neutrinos have non-zero mass and (ii) Dark Matter (DM) makes about 25\% of the total energy density of the Universe. Standard Model (SM) of particle physics, though extremely successful in explaining interactions of elementary particles (except gravity), does not explain neutrino mass and DM, leading to interesting SM extensions in this direction. In particular, DM candidates with masses around the electroweak scale and interaction cross-sections comparable to those of the electroweak interactions can naturally account for the observed DM relic abundance. This intriguing alignment between theoretical expectations and observational evidence is often referred to as the WIMP (Weakly Interacting Massive Particle) Miracle \cite{Hut:1977zn,Jungman:1995df,Bertone:2004pz}. It highlights, how particles with electroweak-scale properties provide a compelling and elegant explanation for DM, linking cosmology to particle physics. The WIMPs are produced thermally in the early universe, where their annihilation rate into SM particles balances the expansion of the Universe, leaving a relic density consistent with current observations when the expansion rate exceeds the annihilation rate, known as the \textit{freeze-out} mechanism.

\noindent One of the most promising ways to generate tiny neutrino mass is the famous seesaw mechanism \cite{Minkowski:1977sc,Yanagida:1979as,Glashow:1979nm,Gell-Mann:1979vob,Mohapatra:1979ia}. However, the Scotogenic model, an extension of the SM, provides a unified framework in which neutrino mass is generated radiatively through quantum loop diagrams and contains natural candidates for WIMP DM \cite{Tao:1996vb,Ma:2006km,Ma:2008cu,Batra:2022pej}. This interplay is crucial in understanding new physics beyond SM relevant for explaining two important unresolved issues: non-zero neutrino mass and DM. The inert doublet in the loop does not acquire vacuum expectation value ($vev$) and is odd under $\mathcal{Z}_2$ symmetry making it a suitable DM candidate. Motivated by the requirement of producing consistent values of relic density of DM, baryon asymmetry of the Universe (BAU), and bounds on charged lepton flavor violation (cLFV), the minimal scotogenic model has been extended in a variety of ways \cite{Kubo:2006yx,AristizabalSierra:2008cnr,Suematsu:2009ww,Hirsch:2013ola,Bonilla:2016diq,Ahriche:2017iar,Cai:2017jrq,Bonilla:2018ynb,Ahriche:2018ger,Avila:2019hhv,CentellesChulia:2019gic,CentellesChulia:2022vpz,Tapender:2024ktc,Kumar:2024zfb,CentellesChulia:2024iom}. Apart from these extensions, another possibility is that of the singlet-triplet scotogenic model \cite{Hirsch:2013ola,Batra:2022org,Singh:2023eye}.

\noindent In this work, we study an extension of the Type-III scotogenic model \cite{Ma:2008cu} incorporating a real scalar singlet from the viewpoint of cosmic inflation \cite{Guth:1980zm}, an idea primarily proposed to deal with Horizon \cite{Rindler:1956yx} and flatness problem \cite{Bondi:1948qk}. Also, the observations of the Cosmic Microwave Background (CMB) \cite{Planck:2018vyg,Planck:2013jfk,BICEP2:2015nss} reveal that the Universe experienced a period of rapid, exponential expansion known as Cosmic Inflation before entering the hot Big Bang phase. This inflationary period can be modelled using scalar fields within the SM that couple non-minimally to the Ricci scalar $R$, enabling accelerated expansion. A prominent scenario in this context is Higgs inflation, which has gained significant interest, especially in light of recent Planck data suggesting its compatibility with cosmological observations. However, challenges arise in Higgs inflation models. If the Higgs scalar doublet serves as the inflaton, vacuum stability and unitarity issues may emerge, as scattering amplitudes among scalars with non-minimal couplings to $R$ could lead to unitarity violation at scales lower than the inflation scale. These issues, as suggested in Ref. \cite{Sher:1988mj,Barbon:2009ya,Hertzberg:2010dc,Kaiser:2010ps,Lerner:2011it}, can be addressed and potentially avoided. There exists a plethora of inflation models (for review, see Ref. \cite{Linde:2014nna}). In our model, the singlet scalar induces inflation similar to the Higgs inflation. The requirement of perturbative unitarity, also, imposes an upper bound on the non-minimal coupling of scalar singlet with Ricci scalar $i.e.$ $\xi < 10^{3}$ \cite{Lebedev:2023zgw}. The inflation in the scotogenic model has been studied in Ref. \cite{Borah:2018rca}. The inflation by singlet scalar in the scotogenic model has been studied in Ref. \cite{Suematsu:2019kst,Hashimoto:2020xoz}. After inflation, the inflaton field decays to the model's particles and reheat the Universe. We consider the mass of the inflaton to be above the electroweak scale, as a result, the inflaton decays completely before DM freeze-out.\\
\noindent Both fermionic and scalar possibilities for dark matter are present in the model. The triplet fermion DM remains under-abundant below 2.5 TeV due to the large annihilation cross-section, however, if we consider coannhilation effects due to scalar doublet in the model, the relic density can be satisfied in the lower mass range of about 1.4 TeV. 

\noindent The model exhibits rich collider phenomenology. Dedicated searches for the triplet fermion in Type-III seesaw have been performed at the LHC~\cite{CMS:2017ybg,CMS:2019lwf,ATLAS:2020wop,ATLAS:2022yhd}. These bounds, however, are not directly applicable to our framework. In our analysis, we have evaluated the mass bound on the triplet fermion within the present model, making use of the existing LHC searches. We find that the CMS and ATLAS results imply a lower bound on the mass of the triplet fermion of about $580~\text{GeV}$~\cite{ATLAS:2022rme,Belanger:2022gqc,Biswas:2023azl}. At higher mass ranges, the production cross-section of the fermion triplet
becomes too small to be effectively probed at the LHC. The $e^{+}e^-$ colliders offer a cleaner environment and enhanced sensitivity for precision measurements, making them a promising avenue for investigating the properties of dark sector particles and interactions in scenarios where the LHC lacks sufficient reach. Also, the combination of beam polarization and control over longitudinal dynamics positions lepton colliders as complementary tools to hadron colliders. The inert doublet, lacking QCD interactions, can be probed at the International Linear Collider (ILC) \cite{ILC:2013jhg} and the Compact Linear Collider (CLIC), a high-luminosity $e^{+}e^-$ collider operating at 380 GeV, 1.5 TeV and 3 TeV \cite{Aicheler:2018arh}. The collider phenomenology of the inert doublet has been studied at $e^{+}e^-$  colliders in Refs. \cite{Kalinowski:2018ylg, Kalinowski:2018kdn} (for review, see Ref. \cite{Belyaev:2021ngh}), while signatures of the fermion triplet at colliders have been explored in Ref. \cite{vonderPahlen:2016cbw,Das:2020uer,Das:2020gnt}. In the present work, we focus on exploring the potential to probe the triplet fermion at both $e^{+}e^-$ (ILC and CLIC) and $pp$ (FCC-hh) colliders with scalar doublet as DM candidate.

\noindent The analyses performed in this paper are organized as follows. In Section \ref{model}, we discuss the extension of the Type-III scotogenic model and mass term for dark sector particles and inflaton field. A scenario, within the present work, has been discussed, in Section \ref{inflation}, to explain the role of scalar singlet as inflaton. The one-loop generation of neutrino mass has been discussed in Section \ref{numass}. The phenomenology of the DM sector, considering two possibilities, such as scalar doublet and fermion triplet DM in the model, is discussed in Section \ref{sec:DM}. Section \ref{collider} deals with the collider signature of the model. Finally, the main conclusions of the paper are summarized in Section \ref{conclusion}. 
\section{Model Set-up}\label{model}
\noindent The particle content of our model and their corresponding charge assignments are given in Table \ref{tab:model}. The SM particle content is extended by two generations of triplet fermions $\Sigma$, an inert scalar doublet $\eta$ and a real singlet scalar $\chi$. This particle content leads to mass generation of light majorana neutrinos at the one loop level. All internal particles in the loop transform as odd, while the SM particles transform as even under the $\mathcal{Z}_2$ symmetry. Consequently, any possible effective operator leading to a decay of a particle of the dark sector necessarily implies another dark sector particle. Therefore, the lightest of them is completely stable and is the DM candidate.

\begin{table}[th]
\begin{center}
\begin{tabular}{ | c | c | c |}
  \hline \hline
& \hspace{0.1cm}  Fields  \hspace{0.1cm}          &  \hspace{0.1cm}  $SU(2)_L \otimes U(1)_Y \otimes \mathcal{Z}_2$               \\
\hline
\multirow{4}{*}{\vspace{0.3cm} \begin{turn}{90} Fermions \end{turn} } &
 $L_i$        	  &    ($\mathbf{2}, {-1/2}, {+}$)    \\
 &   $\ell_{R_j}$       &   ($\mathbf{1}, {-1}, {+}$)     \\
&   $\Sigma$       &   ($\mathbf{3}, {0}, {-}$)     \\
\hline 
\multirow{4}{*}{ \begin{turn}{90} \hspace{0.2cm} Scalars \end{turn} } &
 $\Phi$  		 &  ($\mathbf{2}, {1/2}, {+}$)      \\
& $\eta$          	 &  ($\mathbf{2}, {1/2}, {-}$)     \\
& $\chi$             &  $(\mathbf{1}, {0}, {+})$  \\	
    \hline
  \end{tabular}
\end{center}
\caption{Particle content and symmetry assignments under $SU(2)_{L}\times U(1)_{Y}\times \mathcal{Z}_2$ gauge symmetry.}
 \label{tab:model}
\end{table}

\noindent The full scalar potential of the model is written as
\begin{eqnarray}\label{oddchi}
\begin{aligned}
V &= \mu_\Phi^2\,\Phi^\dagger\Phi + \mu_\eta^2\,\eta^\dagger\eta + \mu_\chi^2\,\chi^2
    + \lambda_1(\Phi^\dagger\Phi)^2 + \lambda_2(\eta^\dagger\eta)^2
    + \lambda_3(\eta^\dagger\eta)(\Phi^\dagger\Phi) \\
  &\quad + \lambda_4(\eta^\dagger\Phi)(\Phi^\dagger\eta)
    + \frac{\lambda_5}{2}\big[(\eta^\dagger\Phi)^2+(\Phi^\dagger\eta)^2\big]
    + \frac{\lambda_6}{4}\chi^4
    + \frac{\lambda_7}{2}\chi^2\,\Phi^\dagger\Phi
    + \frac{\lambda_8}{2}\chi^2\,\eta^\dagger\eta\\
  &\quad + \; \mu_{\chi_3} \chi^3 + \lambda_{\Phi \chi} \chi (\Phi^\dagger \Phi) + \lambda_{\eta\chi} \chi (\eta^\dagger \eta)  \,.
\end{aligned}
\end{eqnarray}

\noindent In this work, we will take the vev of $\chi$ large and its coupling with other scalar small, such that the field $\chi$ does not drastically affect the EW scale physics. For such large field values $\chi \approx \mathcal{O}( 10^{15}\hspace{0.1cm} \text{GeV})$, the dimensionless quartic interactions $\frac{\lambda_6}{4}\chi^{4}$ plays the dominant role in inflation which we will discuss in Sec.~\ref{inflation}. This is standard assumption often taken in singlet inflation models \cite{Lebedev:2012zw,Suematsu:2019kst,Hashimoto:2020xoz,McDonald:2016cdh,Lebedev:2023zgw} under which the scalar potential (Eqn. (\ref{oddchi})) reduces to
\begin{eqnarray} \label{potential}
\begin{aligned}
V =  &\mu_\Phi^2 \Phi^{\dagger} \Phi+\mu_\eta^2 \eta^{\dagger} \eta+\mu_\chi^2 \chi^2+\lambda_1\left(\Phi^{\dagger} \Phi\right)^2+\lambda_2\left(\eta^{\dagger} \eta\right)^2+\lambda_3\left(\eta^{\dagger} \eta\right)\left(\Phi^{\dagger} \Phi\right) \\
& +\lambda_4\left(\eta^{\dagger} \Phi\right)\left(\Phi^{\dagger} \eta\right)+\frac{\lambda_5}{2}\left[\left(\eta^{\dagger} \Phi\right)^2+\left(\Phi^{\dagger} \eta\right)^2\right]+\frac{\lambda_6}{4}\chi^4 +\frac{\lambda_7}{2}\chi^2\Phi^{\dagger} \Phi+\frac{\lambda_8}{2}\chi^2\eta^{\dagger} \eta.
\end{aligned}
\end{eqnarray}

\noindent The scalar fields $\Phi$, $\eta$ and $\chi$ denote the usual SM Higgs scalar doublet, the inert scalar doublet and real\footnote{Here, for sake of simplicity, we are taking the scalar singlet as real.} scalar singlet, respectively.
Expanding the $SU(2)_L$ components of the scalar fields, we can express them as
\begin{eqnarray}
& & \begin{aligned}
  \Phi = \frac{1}{\sqrt{2}}\begin{pmatrix}
         \sqrt{2}G^{+}\\
        v_{\Phi} + h + i G^{0}
     \end{pmatrix}, & \hspace{0.5cm}& \eta = \frac{1}{\sqrt{2}}\begin{pmatrix}
\sqrt{2}\eta^+\\
\eta_{R}+i\eta_{I}
\end{pmatrix}.
\end{aligned}
\end{eqnarray}

\noindent Here $h$ is the SM Higgs and $G^{+}$/$G^{0}$ are the charged/neutral Goldstones. Also $\eta^+$ represents the charged component, while $\eta_R$ and $\eta_I$ correspond to the neutral CP-even and CP-odd components of the inert scalar doublet $\eta$, respectively.
After the Higgs takes the $vev$ $(v_{\Phi})$ and spontaneously breaks the electroweak symmetry, the SM Higgs $(h)$ and Singlet scalar $(\chi)$ mix through the rotation matrix $O_R$ as follows
\begin{eqnarray}
\binom{h_1}{h_2}=O_R\binom{h}{\chi} \equiv\left(\begin{array}{cc}
\cos \alpha & \sin \alpha \\
-\sin \alpha & \cos \alpha
\end{array}\right)\binom{h}{\chi},
\end{eqnarray}
where $\alpha$ is the mixing angle between them.
Then, we have that
\begin{eqnarray}
O_R M_R^2 O_R^T=\operatorname{diag}\left(m_{h_1}^2, m_{h_2}^2\right),
\end{eqnarray}
$M_R^2$ is the squared CP-even mass matrix whose eigenvalues are given by
\begin{eqnarray}\label{mh12}
m_{\left(h_1, h_2\right)}^2=\left(\lambda_1 v_{\Phi}^2+\lambda_6 v_\chi^2\right) \mp \sqrt{\lambda_7^2 v_{\Phi}^2 v_\chi^2+\left(\lambda_1 v_{\Phi}^2-\lambda_6 v_\chi^2\right)^2},
\end{eqnarray}
where the `$-$' (`+') sign corresponds to $h_1\left(h_2\right)$ and $v_{\chi}$ is the $vev$ of singlet scalar field. Note that $\eta$ does not mix with $\chi$ or $\Phi$ as it has odd $Z_2$ charge. 
Furthermore, the masses of the CP-even and CP-odd components of the inert doublet, $\eta$, turn out to be
\begin{eqnarray}\label{eqn6}
m_{\left(\eta_R, \eta_I\right)}^2=\mu_{\eta}^2+\frac{\lambda_8}{2} v_\chi^2+\frac{\lambda_3+\lambda_4 \pm \lambda_5}{2} v_{\Phi}^2 .
\end{eqnarray}
The mass of the charged scalar field is given by,
\begin{eqnarray}\label{eqn7}
m_{\eta^{\pm}}^2=\mu_{\eta}^2+\frac{\lambda_3}{2} v_\Phi^2+\frac{\lambda_8}{2} v_{\chi}^2.
\end{eqnarray} 

\noindent Note that, the large value of 
$v_\chi$ necessitates choosing a small value for the 
$\lambda_8$ coupling.

\noindent The relevant Yukawa Lagrangian based on the particle content and symmetries (Table \ref{tab:model}) is given by
\begin{eqnarray}\label{yukawa}
-\mathcal{L}_{\mathrm{Y}} \supset Y_{i j}^{\ell} \bar{L}_i \Phi \ell_{R_j}+Y_{i k}^\nu \bar{L}_i \tilde{\eta} \Sigma_k+\frac{1}{2} Y^\Sigma \chi \text{Tr}(\bar{\Sigma}^{c}\Sigma)+\frac{1}{2}m_{\Sigma}\text{Tr}(\bar{\Sigma}^{c}\Sigma)+\text { h.c.},
\end{eqnarray}
where $\tilde{\eta} = i \tau_{2} \eta^{*}$, $\tau_{2}$ is the $2^{nd}$ Pauli matrix, $L_i=\left(\nu_{L_i}, \ell_{L_i}\right)^T$ and the indices $i,j \in \{1,2,3\}$, $k \in \{1, 2\}$.

\noindent The triplet fermion can be expressed in its fundamental representation as
\begin{eqnarray}
\Sigma = \begin{pmatrix} \frac{1}{\sqrt{2}} \Sigma^0 & \Sigma^+ \\ \Sigma^- & -\frac{1}{\sqrt{2}} \Sigma^0 \end{pmatrix}.
\end{eqnarray}
Also, the mass of the triplet fermion is given by 
\begin{equation}
\label{eqn10}
M_{\Sigma_{k}}=m_{\Sigma_{k}} + Y^\Sigma \langle \chi \rangle.
\end{equation}
It is important to note that components of the triplet fermion ($\Sigma^{0},\Sigma^{\pm}$) remain degenerate at the tree level. However, a mass splitting arises at the one-loop level.
Throughout the paper, whenever the mass splitting among components of $\Sigma_k$ is not important,  we will use $M_{\Sigma_k}$ to collectively denote the masses of the quasi-degenerate components of $\Sigma_k$.
Furthermore, we define $M_{\Sigma^0_{k}}$ as the mass of the neutral component of the triplet fermion $\Sigma^0_k$ and $M_{\Sigma^{\pm}_{k}}$ as the mass of its charged component $\Sigma^{\pm}_k$.

In the next section, we will delve into the inflationary dynamics driven by the singlet scalar field. This analysis will explore how the singlet scalar field couples to the Ricci scalar (in general relativity) and its role in shaping the inflationary potential. We will also discuss the slow-roll conditions, calculate the associated inflationary observables, and examine the parameter space that leads to successful inflation consistent with current cosmological data.
\section{Inflation}\label{inflation}
\noindent In our model, we take an approach where only the singlet scalar $\chi$ is allowed to couple significantly with the Ricci scalar. The singlet scalar inflation has been studied in Ref. \cite{Lerner:2009xg,Lerner:2011ge,Suematsu:2012gk}. Defining the field $\chi = \langle \chi \rangle + \tilde{\chi}$, where $\langle \chi \rangle$ is the vacuum expectation value and $\tilde{\chi}$ represents quantum fluctuations around $vev$, then $\chi$ can act as the inflaton, providing a viable inflationary mechanism without compromising unitarity at the relevant scales.
We assume that only the field $\chi$ has a significant non-minimal coupling with the Ricci scalar.

\begin{figure}[th]
    \centering
\includegraphics[height=7cm,width=12cm]{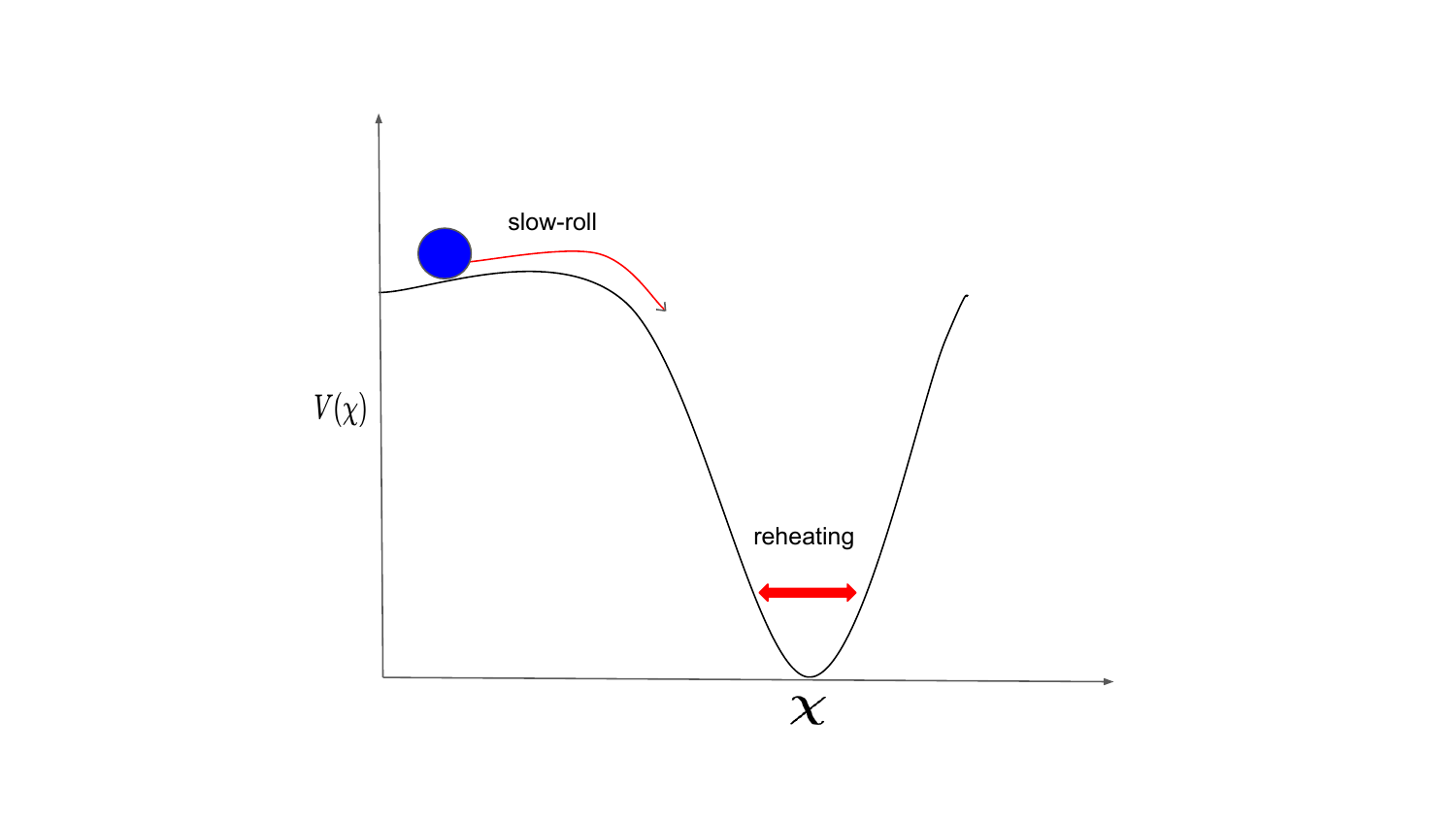}
    \caption{The schematic diagram of slow-roll inflation.}
    \label{slow-roll}
\end{figure}

\noindent In the limit where $v_{\chi}>>v_{\Phi}$, we can use the Binomial expansion to rewrite the Eqn. (\ref{mh12}) as
\begin{eqnarray}
m^{2}_{h_{1}}& \approx & \left(2 \lambda_{1}-\frac{\lambda_{7}^{2}}{\lambda_{6}}\right)v_{\Phi}^{2},\\
m^{2}_{\chi}& \approx & 2\lambda_{6} v^{2}_{\chi}.
\end{eqnarray}
Here, we consider $m_{\chi}$\footnote{We consider $m_{\chi}\sim m_{h_2}$.} as the mass of the inflaton field, which is responsible for driving cosmic inflation in the early Universe and generating the exponential expansion of space during the inflationary epoch, smoothing out initial irregularities and providing the initial conditions for the subsequent evolution of the Universe. In general relativity, the dynamics of gravity in the absence of other fields is described by the Einstein-Hilbert action, given as
\begin{eqnarray}
S_{EH}=\int d^4 x \sqrt{-g}\left[\frac{1}{2}M^2_{P}R\right],
\end{eqnarray}
where $M^2_{P}$ is the reduced Planck mass, $R$ is Ricci scalar which encodes the spacetime curvature.
For the inflationary scenario in the Jordan frame, the relevant action is:
\begin{eqnarray}
\chi_{J} = \int d^4 x \sqrt{-g} \left[ -\frac{1}{2} M_{\mathrm{pl}}^2 R - \frac{1}{2} \xi \chi^2 R + \frac{1}{2} \partial^\mu \chi \partial_\mu \chi - V(\chi) \right],
\label{eqn13}
\end{eqnarray}
where $g$ is the determinant of the metric tensor $g^{\mu\nu}$ and $V(\chi)$ is the potential associated with the inflaton field $\chi$. During inflation, other scalar fields assumed to have negligible values compared to $\chi$. 
For large values of $\chi$, the potential $V(\chi)$ can be approximated as $V(\chi) \simeq \lambda_6 \chi^4$, where $\lambda_6$ is a coupling constant that governs the inflaton self-interaction, making this model of inflation primarily driven by quartic self-interactions of the inflaton field.
Further, to perform the conformal transformation from the Jordan frame to the Einstein frame, we need to rescale the metric such that the gravitational part of the action takes the standard Einstein-Hilbert form. The conformal transformation of the metric given as
\begin{eqnarray}
g_{\mu \nu}=\Omega^2 \tilde{g}_{\mu \nu}, \quad \Omega^2=1+\frac{\xi \chi^2}{M_{\mathrm{pl}}^2}.
\end{eqnarray}
As a result of the conformal transformation, the action in the Einstein frame can be written as
\begin{eqnarray}
\begin{aligned}
\chi_{E}= & \int d^4 x \sqrt{-\tilde{g}}\left\{-\frac{1}{2} M_{\mathrm{pl}}^2 R_E+\frac{1}{2 \Omega^4}\left[1+\frac{\left(\xi+6 \xi^2\right) \chi^2}{M_{\mathrm{pl}}^2}\right]\right. \\
& \left.\times \partial^\mu \chi \partial_\mu \chi-\frac{1}{\Omega^4} V(\chi)\right\}.
\end{aligned}
\label{eqn14}
\end{eqnarray}
The action in Eqn. (\ref{eqn14}) can be expressed by a canonically normalized field $\gamma$ as
\begin{eqnarray}
\frac{d \gamma}{d \chi}=\frac{\left[1+\left(\xi+6 \xi^2\right) \frac{\chi^2}{M_{\mathrm{pl}}^2}\right]^{1 / 2}}{1+\frac{\xi \chi^2}{M_{\mathrm{pl}}^2}}.
\end{eqnarray}
In the region with condition $\chi<<M_{pl}/\sqrt{\xi}$ satisfied, the canonical normalized field $\gamma$ coincides with $\chi$. Alternately, in the region where $\chi>>M_{pl}/\sqrt{\xi}$ is satisfied, the potential has the following form
\begin{eqnarray}
V_{SR}=\frac{\lambda_{6}\chi^{4}}{4\left(1+\frac{\xi\chi^{2}}{M^{2}_{pl}}\right)^{2}}.
\end{eqnarray}
This implies that in this region, $\chi$ may act as the slow-rolling inflation field. During this period, the potential energy $V_{\chi}$ of the inflaton field ($\chi$) dominates over its kinetic energy, ensuring a quasi-exponential expansion of the universe. The slow-roll conditions are critical to ensuring that inflation lasts long enough (typically $50-60$ $e$-foldings) and ends smoothly, as shown in Fig.~\ref{slow-roll}.\\
From potential $V_{SR}$, the number of $e$-foldings can be estimated as
\begin{eqnarray}
N= \int_{\gamma}^{\gamma_{end}} \frac{V_{SR}}{M^{2}_{pl} 
 dV_{SR}/d\gamma} \,d\gamma \simeq \frac{3}{4}\frac{(\gamma^2-\gamma^2_{end})\xi}{M^{2}_{pl}}.
 \label{eqn18}
\end{eqnarray}
 The slow-roll parameters\footnote{$\epsilon$ quantifies how slowly the inflaton is rolling down its potential while $\eta$ measures the curvature of the potential and the acceleration of the inflaton} ($\epsilon,\eta$) can be calculated from the potential $V_{SR}$ as
\begin{eqnarray}
\epsilon=\frac{M^{2}_{pl}}{2}\frac{dV_{SR}/d\gamma}{V_{SR}}\simeq \frac{4M^{4}_{pl}}{3\xi^{2}\chi^{4}}, \hspace{1cm} \eta=M^{2}_{pl}\frac{d^2V_{SR}/d\gamma^2}{V_{SR}}\simeq -\frac{4M^{2}_{pl}}{3\xi\chi^{2}}.
\end{eqnarray}
For the successful inflation, the slow-roll parameters $\epsilon,\eta<<1$. The inflation will end at $\epsilon\simeq1$, therefore, at this point $\gamma_{end}\simeq \sqrt{4/3}M_{pl}^{2}/\xi$, which can be neglected in Eqn. (\ref{eqn18}). Therefore the slow-roll parameters can be expressed in terms of the number of $e$-folding parameter $N$ as 
\begin{eqnarray}
\epsilon\simeq\frac{3}{4N^{2}}, \hspace{1cm} \eta\simeq -\frac{1}{N}.
\end{eqnarray}
The spectral index $n_{s}$ and tensor-to-scalar ratio $r$ are represented by using the slow-roll parameters as
\begin{eqnarray}
n_{s}=1-6\epsilon+2\eta, \hspace{1cm} r=16\epsilon.
\end{eqnarray}
Also, the primordial power spectrum, which describes the distribution of initial density fluctuations in the early Universe, typically arising from quantum fluctuations during cosmic inflation, is given by
\begin{eqnarray}
\mathcal{P}(k)=A_{s}\left(\frac{k}{k_{*}}\right)^{n_{s}-1} \hspace{0.5cm} \text{with} \hspace{0.5cm} A_s=\frac{V_{SR}}{24 \pi^{2}M^{2}_{pl}\epsilon}\Big|_{k_{*}},
\end{eqnarray}
is amplitude of the spectrum at the pivot scale $k_{*}$ and $k$ is the spatial scale or wave number. By using Planck data, $A_{s}=(2.101^{+0.031}_{-0.034})\times10^{-9}$ at $k_{*}=0.05Mpc^{-1}$ \cite{Planck:2018vyg}, we find the relation
\begin{eqnarray}\label{lam6}
\lambda_{6}\simeq1.49\times10^{-6}\xi^{2}N^{-2}.
\end{eqnarray}
For $N=60$, the slow-roll parameters yields $n_s \sim 0.965$ and $r \sim 3.3 \times 10^{-3}$.
 The experimental value of $n_s$ from Planck data is $n_s = 0.9649 \pm 0.0042$, with an upper bound on $r < 0.036$ set by combined data from BICEP, Keck Array, and Planck \cite{BICEP2:2018kqh,Planck:2018vyg}. The requirement of perturbative unitarity imposes an upper bound on the non-minimal coupling, i.e., $\xi < 10^{3}$ \cite{Lebedev:2023zgw}. Consequently, from Eqn. (\ref{lam6}), $\lambda_{6}$ cannot exceed 0.41 for the maximum value of $\xi$ at an $e$-folding number of $N = 60$ to achieve the observed value of $A_{s}$. To ensure consistency, choosing $\xi = \mathcal{O}(10^2)$ reproduces the observed amplitude of the spectrum, $A_s$, for $\lambda_6 = \mathcal{O}(10^{-6})$. 
 
\noindent Inflation concludes when the potential stabilizes at a minimum or equilibrium point. At this stage, the inflaton field starts oscillating around the minima of the potential, resembling a harmonic oscillator or the vacuum \( \langle \chi \rangle \). These oscillations lead to the inflaton decaying into other particles within the model. This marks the onset of the reheating phase, a critical period where the Universe repopulated with particles and radiations. During reheating, energy is transferred from the inflaton field to the model particles, initiating the hot Big Bang and setting the stage for the subsequent evolution of the Universe. Within the model, the inflaton field decays into an inert doublet and a fermion triplet at the tree level, and it can further decay into SM particles such as gauge bosons, neutrinos and Higgs through one-loop processes, where $\eta$ and $\Sigma$ appear in the internal propagator lines, as depicted in Fig.~\ref{dc}. This decay mechanism provides a pathway for the inflaton to transfer its energy to SM particles, possibly influencing the reheating process and the subsequent evolution of the early Universe. The decay width could be estimated as\footnote{See Ref. \cite{Hashimoto:2020xoz} for analytical calculations.}
\begin{figure}[t]
    \centering
   \begin{tikzpicture}
\begin{feynman}
\vertex at (0,0) (i1);
\vertex at (-2,0) (i2);
\vertex at (1,1) (a);
\vertex at (1,-1) (b);
\diagram*{
(i2) -- [scalar, edge label=\(\chi\)] (i1), (i1) -- [scalar, edge label=\(\eta\)] (a), (i1) -- [scalar, edge label=\(\eta\)] (b),
};
\end{feynman}
\end{tikzpicture}
\hspace{1cm}
 \begin{tikzpicture}
\begin{feynman}
\vertex at (0,0) (i1);
\vertex at (-2,0) (i2);
\vertex at (1,1) (a);
\vertex at (1,-1) (b);
\diagram*{
(i2) -- [scalar, edge label=\(\chi\)] (i1), (i1) -- [fermion, edge label=\(\Sigma\)] (a), (i1) -- [fermion, edge label=\(\Sigma\)] (b),
};
\end{feynman}
\end{tikzpicture}
\hspace{1cm}  \begin{tikzpicture}
\begin{feynman}
\vertex at (-1,0) (i1);
\vertex at (1,-1) (b);
\vertex at (-2.6,0) (c);
\vertex at (-4,0) (d);
\diagram*{
(i1) -- [photon, edge label=\(G_{\mu}\)] (a), (i1) -- [photon, edge label=\(G_{\nu}\)] (b), (c) -- [scalar, edge label=\(\chi\)] (d),
};
\draw[dashed,decoration={markings, mark=at position 0.5 with {\node[above] {\(\eta\)};}}, postaction={decorate}] (-1,0) arc (0:180:0.8); 

\draw[dashed, decoration={markings, mark=at position 0.5 with {\node[below] {\(\eta\)};}}, postaction={decorate}] (-1,0) arc (360:180:0.8); 
\end{feynman}
\end{tikzpicture} 
\hspace{1cm}
\begin{tikzpicture}
\begin{feynman}
\vertex at (-1,0) (i1);
\vertex at (1,-1) (b);
\vertex at (-2.6,0) (c);
\vertex at (-4,0) (d);
\diagram*{
(i1) -- [photon, edge label=\(G_{\mu}\)] (a), (i1) -- [photon, edge label=\(G_{\nu}\)] (b), (c) -- [scalar, edge label=\(\chi\)] (d),
};
\draw[decoration={markings, mark=at position 0.5 with {\node[above] {\(\Sigma\)};}}, postaction={decorate}] (-1,0) arc (0:180:0.8); 

\draw[decoration={markings, mark=at position 0.5 with {\node[below] {\(\Sigma\)};}}, postaction={decorate}] (-1,0) arc (360:180:0.8); 
\end{feynman}
\end{tikzpicture} 
\caption{The decay modes of the inflaton field at tree and one-loop levels are considered. Here, $G_{\nu}$ represents $W^{\pm}$ and $Z$ gauge bosons. The SM Higgs boson ($h$) is, also, permitted as a possible final state.}
\label{dc}
\end{figure}
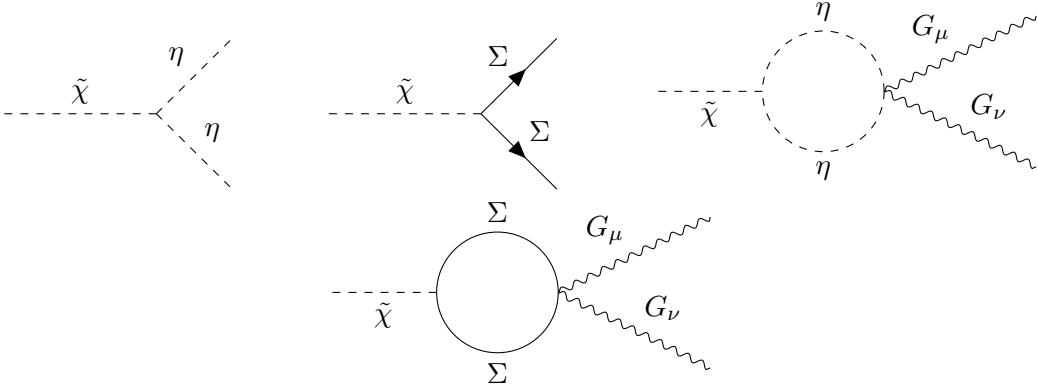
\begin{eqnarray}\nonumber
&\Gamma_{\tilde{\chi}} &= \frac{(\lambda_8 \langle \chi \rangle)^2}{32 \pi m_{\chi}} \sqrt{1 - \frac{4 M_\eta^2}{m_{\chi}^2}} + \frac{Y^{\Sigma} m_{\chi}}{8 \pi} \left( 1 - \frac{4 M_\Sigma^2}{m_{\chi}^2} \right)^{3/2}\\
&+& \frac{(\lambda_8 \langle \chi \rangle)^2}{4096 \pi^5 m_{\chi}} \left[ \frac{\left(2 c_w^4 + 1\right) g^4}{c_w^4} \left| \mathcal{I}\left( \frac{M_{\eta/\Sigma}^2}{m_{\chi}^2} \right) \right|^2 + \frac{1}{2} (4\lambda_3 + 2\lambda_4)^2 \left| \mathcal{J}\left( \frac{M_{\eta/\Sigma}^2}{m_{\chi}^2} \right) \right|^2 \right],
\label{gammachi}
\end{eqnarray}
where
\begin{eqnarray}
& \mathcal{I}(x)=1+x\left(\ln \frac{1+\sqrt{1-4 x}}{1-\sqrt{1-4 x}}+i \pi\right)^2, \\
& \mathcal{J}(x)=\sqrt{1-4 x}\left(\ln \frac{1+\sqrt{1-4 x}}{1-\sqrt{1-4 x}}+i \pi\right)-2.
\end{eqnarray}
Here, $g$ denotes the $SU(2)$ gauge coupling, and $c_{w}$ represents the cosine of the weak mixing angle. Also, $M_{\eta}$ and $M_{\Sigma_k}$ denote the masses of the neutral components of inert doublet scalar (Eqn.~\ref{eqn6}) and triplet fermions (Eqn.~\ref{eqn10}), respectively.  Additionally, the term $\sqrt{1 - \frac{4 M_{\eta (\Sigma)}^2}{m_{\chi}^2}}$ is a kinematic factor, accounting for the phase space suppression during the inflaton's decay to the $\eta$ ($\Sigma$) states. In the scenario where the inflaton field is significantly more massive than other fields in the model, this kinematic suppression factor approaches unity, effectively removing phase space constraints and allowing for efficient decay into these particles. Therefore, the Eqn. (\ref{gammachi}) becomes
\begin{eqnarray}
\Gamma_{\tilde{\chi}}= \frac{(\lambda_8 \langle \chi \rangle)^2}{32 \pi m_{\chi}} + \frac{Y_{\Sigma}^2 m_{\chi}}{8 \pi}+ \frac{(\lambda_8 \langle \chi \rangle)^2}{4096 \pi^5 m_{\chi}} \left[ \frac{\left(2 c_w^4 + 1\right) g^4}{c_w^4}+ \frac{1}{2} (4\lambda_3 + 2\lambda_4)^2\right].
\end{eqnarray}
At $\Gamma_{\tilde{\chi}}=H$, the reheating temperature after the inflation is given by 
\begin{eqnarray}
T_{R}=\left(\frac{90}{\pi^{2}g_{*}}\right)^{1/4}\sqrt{\Gamma_{\chi}M_{pl}},
\end{eqnarray}
where $g_{*}=121.75$ is the relativistic degree of freedom.

\noindent At the onset of inflaton oscillations, energy loss to particles is minimal compared to that from Hubble expansion rate ($H$). Only when the Hubble rate drops to about \(\Gamma\) does actual reheating become significant.
We numerically computed the inflation parameters for the number of $e-$ folding $N = 60$, varying $\lambda_6$ randomly over the range from $10^{-6}$ to 1. The benchmark values of various parameters such as $\xi$, $\lambda_{6}$, $\lambda_{8}$, $m_{\chi}$, $Y^{\Sigma}$ and reheating temperature $T_{R}$ is given in the Table \ref{bp}. We obtained the reheating temperature $T_{R}\approx\mathcal{O}(10^{10})$ GeV.
\begin{table}[th]
    \centering
    \begin{tabular}{|c |c |c |c |c |c |c |c |}
    \hline
    \hline
 $m_{\chi}(\mathrm{GeV})$ & $\lambda_6$ & $\lambda_8$ & $\xi$ & $Y^{\Sigma}$ & $T_R(\mathrm{GeV})$ \\
\hline\hline
 $8.45\times 10^{6}$ & $3.57\times10^{-6}$ & $1.75\times10^{-6}$  & 29.38 & $8.20\times10^{-6}$ & $2.35 \times 10^{10}$\\
$2.09\times 10^{6}$ & $2.19\times10^{-6}$ & $3.13\times10^{-5}$ & 72.77 & $9.56\times10^{-6}$ & $8.02 \times 10^{10}$ \\
 $2.86 \times 10^{6}$ & $4.08\times10^{-6}$ & $5.16\times10^{-6}$ &  99.24 & $3.73\times10^{-6}$ & $7.12 \times 10^{10}$  \\
 $2.74 \times 10^{6}$ & $3.75\times10^{-6}$ & $4.11\times10^{-5}$ &  95.16 & $7.48\times10^{-6}$ & $1.68 \times 10^{10}$  \\
 $2.04 \times 10^{6}$ & $2.07\times10^{-6}$ & $3.16\times10^{-6}$ &  70.79 & $9.80\times10^{-6}$ & $1.19 \times 10^{11}$\\
\hline
\end{tabular}
\caption{Five representative values of parameters and corresponding reheating temperature $T_R$.}
\label{bp}
\end{table}
\noindent We assume that the $vev$ of the singlet scalar is extremely larger than the electroweak scale. This leads to the inflaton field acquiring a mass much higher than any other particles in the model, ensuring that it decays completely into lighter particles, including SM particles. As a result, no relics of the inflaton remain after the reheating phase, and the $\chi$ field becomes fully decoupled from the rest of the model's particle interactions. This decoupling isolates $\chi$ from the post-reheating particle dynamics, allowing the model to evolve without additional influences from the inflaton.

\section{Neutrino Mass}\label{numass}
\noindent In the model under consideration, the $\mathcal{Z}_2$ symmetry forbids neutrino masses at the tree level, however, they can be generated at the one-loop level. The relevant invariant Yukawa Lagrangian for neutrino mass generation is provided in Eqn. (\ref{yukawa}). It is important to note that, the real singlet scalar (inflaton) remains effectively decoupled from the electroweak sector due to its large mass \(\mathcal{O}(10^{8}~\mathrm{GeV})\) and highly suppressed couplings 
\((\lambda_7, \lambda_8 \sim 10^{-6})\). As a result, it does not affect neutrino mass generation or low-energy lepton-sector phenomenology, while still playing a role in the early-Universe dynamics as discussed earlier. Therefore, the $Y^\Sigma \chi \text{Tr}(\bar{\Sigma}^{c}\Sigma)$ term in Eqn. (\ref{yukawa}) does not contribute to the neutrino mass, as the $\chi$ field decouples from the other model particles after the reheating process.
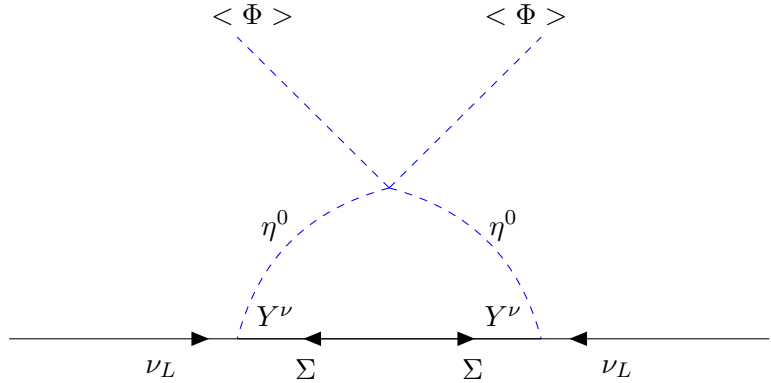
\begin{figure}[th]
    \centering
   \begin{tikzpicture}
\begin{feynman}
\vertex at (2,0) (i1);
\vertex at (-2,0) (i2);
\vertex at (0,0) (a);
\vertex at (0, 2) (d);
\vertex at (5,0) (b);
\vertex at (-5,0) (c);
\vertex at (1.1,-0.4) () {\(\Sigma\)};
\vertex at (-1.1,-0.4) () {\(\Sigma\)};
\vertex at (2,4) (f);
\vertex at (-2,4) (g);
\vertex at (1.8,4.3) () {\(<\Phi>\)};
\vertex at (-1.8,4.3) () {\(<\Phi>\)};
\vertex at (1.5,1.5) () {\(\eta^{0}\)};
\vertex at (-1.5,1.5) () {\(\eta^{0}\)};
\vertex at (-3,-0.4) () {\(\nu_{L}\)};
\vertex at (3,-0.4) () {\(\nu_{L}\)};
\vertex at (1.5,0.3) () {\(Y^{\nu}\)};
\vertex at (-1.5,0.3) () {\(Y^{\nu}\)};
\diagram*{
(a) -- [fermion] (i1), (a) -- [fermion] (i2),
(b) -- [fermion] (a), (c) -- [fermion] (a),
(d) -- [blue, scalar, bend left] (i1), (d) -- [blue, scalar, bend right]  (i2), (d) -- [blue, scalar] (f),
 (d) -- [blue, scalar] (g), 
};
\end{feynman}
\end{tikzpicture}
    \caption{The Feynman diagram used to generate neutrino masses at one-loop in Type-III scotogenic model.}
    \label{fig:neutrinomass}
\end{figure}
Consequently, neutrino masses arise similar to the minimal Type-III scotogenic model \cite{Ma:2008cu}, enabling the model to accommodate neutrino masses without additional contributions from $\chi$ post-reheating. The Feynman diagram illustrating the one-loop mechanism for neutrino mass generation is shown in Fig.~\ref{fig:neutrinomass}, and the corresponding equation is given by
\begin{eqnarray}
\left(M_\nu\right)_{i j}&=&\sum_{k=1}^2 \frac{Y_{i k}^\nu Y_{k j}^\nu M_{\Sigma^0_k}}{32 \pi^2}\left[\frac{m_{\eta_R}^2}{m_{\eta_R}^2-M_{\Sigma^0_k}^2} \log \frac{m_{\eta_R}^2}{M_{\Sigma^0_k}^2}-\frac{m_{\eta_I}^2}{m_{\eta_I}^2-M_{\Sigma^0_k}^2} \log \frac{m_{\eta_I}^2}{M_{\Sigma^0_k}^2}\right],
\end{eqnarray}
where $M_{\Sigma^0_{k}}$ ($k=1,2$) is the mass of neutral components of triplet fermion\footnote{For simplicity, we assume $m_{\Sigma_{k}}\sim M_{\Sigma_{k}}$ after inflationary dynamics.}

\noindent
The neutrino mass matrix $\left(M_\nu\right)_{i j}$ can also be expressed as
\begin{equation*} 
(M_{\nu_{i j}})=(Y^\nu)^T\Lambda Y^\nu 
\end{equation*}
\\
where $\Lambda$ is given by
\begin{equation*}
    \Lambda=\frac{M_{\Sigma^0_k}}{32\pi^2}\left[\frac{m_{\eta_R}^2}{m_{\eta_R}^2-M_{\Sigma^0_k}^2} \log \frac{m_{\eta_R}^2}{M_{\Sigma^0_k}^2}-\frac{m_{\eta_I}^2}{m_{\eta_I}^2-M_{\Sigma^0_k}^2} \log \frac{m_{\eta_I}^2}{M_{\Sigma^0_k}^2}\right]
\end{equation*}

Using the Casas-Ibarra parametrization \cite{Casas:2001sr,Ibarra:2003up}, the Yukawa terms can be written as

\begin{equation*}
    Y^\nu=\sqrt{\Lambda^{-1}}R\sqrt{m_\nu} U_{lep}^{\dagger}
\end{equation*}
where $U_{lep}$ is the leptonic mixing matrix, $m_\nu$ is the diagonal matrix of neutrino mass eigenvalues and R is a complex rotation matrix.

\noindent In the following section, we explore the dark matter phenomenology, constraining the model’s parameter space by utilizing neutrino oscillation data within the 3$\sigma$ experimental range for normal hierarchy\footnote{The results for the inverted hierarchy remain the same.} only \cite{ParticleDataGroup:2024cfk}.
\section{Constraints on the Model}
\noindent The numerical estimation of various parameters is put to the following theoretical and experimental constraints.

\subsection{Boundedness from below}
\noindent Bounded from below for the full scalar potential (Eqn.~\ref{potential}), is ensured by the vacuum stability constraints \cite{Bonilla:2019ipe,DeRomeri:2022cem} such as

\begin{equation}
\lambda_1, \lambda_2, \lambda_6  \geq 0,
\label{eqn:32}
\end{equation}
\begin{equation}
\lambda_3  \geq-2\sqrt{\lambda_1 \lambda_2},
\label{eqn:33}
\end{equation}
\begin{equation}
\lambda_7  \geq-2\sqrt{\lambda_1 \lambda_6},
\label{eqn:34}
\end{equation}
\begin{equation}
\lambda_8  \geq-2\sqrt{\lambda_2 \lambda_6},
\label{eqn:35}
\end{equation}
\begin{equation}
\lambda_3+\lambda_4-\left|\lambda_5\right|  \geq-2\sqrt{\lambda_1 \lambda_2} .
\label{eqn:36}
\end{equation}

\noindent In our scenario, however, $\chi$ obtains a high-scale $vev$ (inflation scale), and the associated physical excitation is heavy and decouples from the electroweak sector. Its couplings to $\Phi$ and $\eta$ ($\lambda_7$ and $\lambda_8$) are taken to be very small ($\sim 10^{-6}$) to suppress $\chi$--Higgs mixing and to ensure inflationary decoupling. After $\chi$ settles to its $vev$, the effective low-energy potential relevant for neutrino mass generation and dark matter reduces to the inert doublet potential with vacuum stability constraints given by Eqs.~(\ref{eqn:32},\ref{eqn:33}) and (\ref{eqn:36}).

\FloatBarrier
\subsection{Electroweak precision data}
\noindent The various parameters in the model can be constrained by the data from electroweak precision data. Particularly, it is known that the oblique parameters $S$, $T$, and $U$ represent the influence of heavy new fields on the gauge boson propagators. In numerical analysis, we utilize the bounds \cite{ParticleDataGroup:2024cfk}
\begin{eqnarray}
S &=& -0.04 \pm 0.10, \\
T &=& 0.01 \pm 0.12, \\
U &=& -0.01 \pm 0.09.
\end{eqnarray}

\subsection{Higgs Boson mass}
\noindent Dark matter particles interacting with the Higgs boson contribute to loop corrections that affect the Higgs boson mass \cite{Bharadwaj:2024crt}. Therefore, we are applying the experimental constraint based on the mass of the Higgs boson \cite{ParticleDataGroup:2024cfk}.

\subsection{Neutrino oscillation data}
\noindent We constrain the DM parameter space by using the neutrino oscillation data in the 3$\sigma$ experimental range \cite{ParticleDataGroup:2024cfk}. We obtained the Yukawa parameters by using Casas-Ibarra parameterisation \cite{Casas:2001sr,Ibarra:2003up}.

\subsection{LEP Constraints}
\noindent The precise LEP-I measurements of the Z-boson decay width impose lower bounds on the masses of dark neutral scalar particles. This limit is simply $m_{\eta_R} + m_{\eta_I} > m_Z$ ensures that the decay $Z \rightarrow \eta_{R} \eta_{I}$ is kinematically forbidden. Also, reinterpreting LEP-II results from chargino searches, in the context of singly-charged scalar production, provides a conservative bound on the mass of dark charged scalar particles $m_{\eta^+} > 70$ GeV \cite{Belyaev:2016lok,Cao:2007rm}. The decays of gauge bosons into $\mathcal{Z}_{2}$-odd pairs are excluded by their invisible width measurements, leading to the constraints, $M_{\Sigma^0_k} + M_{\Sigma^\pm_k} > M_W,\hspace{.2cm}2 M_{\Sigma^\pm_k} > M_Z $. Further, since the $\mathcal{Z}_{2}$-odd fermions couple to gauge bosons with the same couplings as the gauginos, we can apply the bounds on direct chargino searches at LEP~II, $M_{\Sigma^\pm_k} > 103.5~\text{GeV}$,
see Refs. \cite{ALEPH:2003acj,DELPHI:2003uqw,L3:1999onh, vonderPahlen:2016cbw}.

\subsection{LHC constraints on $\Sigma^{\pm}$}\label{lhc}
\noindent The collider bound on $M_{\Sigma_1}$ arises from a search strategy analogous to that used for supersymmetric charginos. The charged component of the triplet fermion, $\Sigma_1^{\pm}$, predominantly decays $via$ 
$\Sigma_1^{\pm} \to \Sigma_1^{0}\pi^{\pm}$ due to a radiatively induced mass splitting 
$\Delta M = M_{\Sigma_1^{\pm}}-M_{\Sigma_1^{0}} \simeq 166~\text{MeV}$ \cite{Cirelli:2005uq,Belanger:2022gqc,Biswas:2023azl}.
\begin{figure}[h!]
 \centering
        \includegraphics[height=6.5cm]{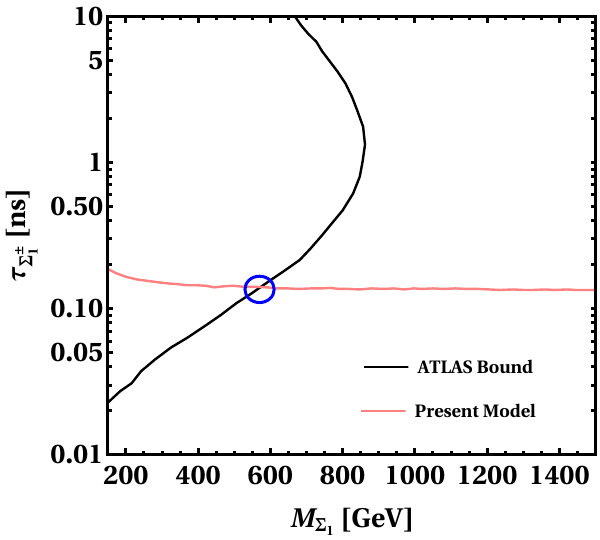}
        \caption{Lifetime of $\Sigma_1^{\pm}$ as a function of the triplet fermion mass, compared with ATLAS limits from disappearing charged track searches.}
        \label{fig:LHC}
\end{figure}
This small gap leads to a significant lifetime $\tau \sim 0.1~\text{ns}$ ($c\tau \approx 3~\text{cm}$), producing the characteristic disappearing-track signature at colliders. The ATLAS constraints on such disappearing charged track signatures \cite{ATLAS:2022rme} for a long-lived chargino decaying into a pion and wino dark matter is shown as the solid black line in Fig. \ref {fig:LHC}. Since the triplet fermion is a multiplet similar to the multiplet containing chargino and wino, we compare our model predictions against this constraint from ATLAS. The lifetime predictions for $\Sigma_1^{\pm}$ as a function of triplet fermion mass is shown as the solid red line in Fig. \ref{fig:LHC}. ATLAS searches for such events with 13~TeV, 136~fb$^{-1}$ data exclude triplet fermion masses below $M_{\Sigma_1} \gtrsim 580~\text{GeV}$ at 95\% CL.

\section{Dark Matter Phenomenology}\label{sec:DM}
\noindent In our model, we have two DM candidates: the real part of an inert scalar doublet ($\eta_{R}$) and the neutral component of the lightest triplet fermion ($\Sigma_{1}^{0}$). In this work, we shall phenomenologically explore both DM scenarios. We set $m_{\chi}$ to be sufficiently large so that the inflaton decays well before DM freeze-out. In order to study the DM phenomenology, we implement the model in the SARAH 4.15.1\cite{Staub:2015kfa} \footnote{See \cite{Vicente:2015zba} for a pedagogical introduction to the use of computational tools in the particle physics such as SARAH, SPheno, micrOMEGAs and MadGraph.} to generate the model files and subsequently employed in the SPheno 4.0.5 version \cite{Porod:2011nf} to calculate various mass matrices and vertices. Finally, in order to numerically solve the Boltzmann equation to calculate the DM relic density and spin-independent DM direct detection cross-section, we implement the model files in the micrOMEGAs-5.3.41\cite{Belanger:2014vza}. We performed a detailed numerical scan with input parameters given in Tab.~\ref{tab:parameterrange}.

\begin{table}[h!]
\begin{center}
\begin{tabular}{  |  c  |     c  |   c | c |}
  \hline\hline 
  Parameter    &   Range   &   Parameter    &   Range  \\
\hline\hline
$\lambda_{1}$     &  	 $[10^{-2},0.128]$            &  $\lambda_{2}$     &  	 $[10^{-8},4\pi]$          \\
$|\lambda_{3}|$   &  $[10^{-8},4\pi]$            &  $|\lambda_{4}|$     &      $[10^{-8},\sqrt{4\pi}]$          \\
$|\lambda_5|$   &   $[10^{-6},\sqrt{4\pi}]$     &  $\mu^2_{\eta}$  &  $[10^{2},10^{8}]\text{ GeV}^2$  \\ 
$m_{\Sigma_1}$   &   $[10,10^4]$\text{ GeV}    &  $m_{\Sigma_2}$  &  $[10,10^4]$\text{ GeV} \\
    \hline
  \end{tabular}
\end{center}
\caption{The ranges of input parameters used in the numerical analysis to calculate DM relic density and direct detection cross-section.}
 \label{tab:parameterrange} 
\end{table}

\subsection{Scalar Doublet Dark Matter ($\eta_{R}$)}
\begin{figure}[t]
 \centering
\hspace{-2cm} \textbf{\Large $\eta_{R}$ as DM}\\
        \includegraphics[height=6.5cm]{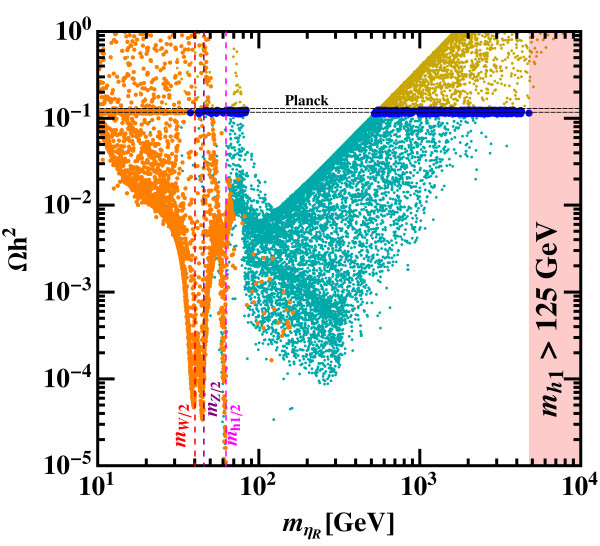}
        \includegraphics[height=6.5cm]{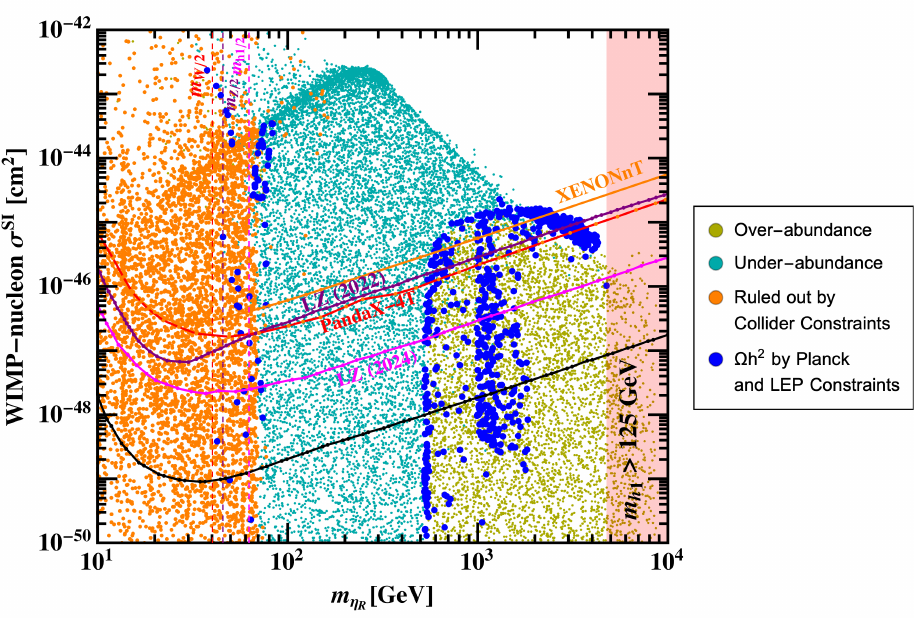}
        \caption{Predictions for the doublet DM case. In both panels, yellow/cyan points represent over/under abundant relic density \cite{Planck:2018vyg}, respectively. The orange-colored points are excluded by LEP constraints \cite{Belyaev:2016lok,Cao:2007rm}. The blue points satisfy the correct relic density. \textbf{Left panel:} Relic density vs mass of the doublet DM particle. \textbf{Right panel:} Spin-independent WIMP-nucleon direct detection cross section vs mass of the doublet DM particle. Red shaded region is ruled out by Higgs constraints similar to \cite{Bharadwaj:2024crt}.}
        \label{fig:doubletDM}
\end{figure}

\noindent We start our DM analysis by focusing on the case in which the lightest state of the dark sector is the neutral component of inert scalar doublet $\eta_R$. The left panel of Fig.~\ref{fig:doubletDM} shows the dependence of the DM relic abundance on the mass of the doublet DM particle, taking into account its annihilation and co-annihilation diagrams into SM particles (Fig.~\ref{fig:relicEta} in Appendix \ref{sec:appendix1}). The parameter space utilized for the numerical analysis is outlined in Table \ref{tab:parameterrange}. We are varying the value of $\lambda_1$ to fix the Higgs boson mass at $m_{h_1} = 125.20 \pm 0.11 $ GeV \cite{ParticleDataGroup:2024cfk}.
We have selected the normal hierarchy for the neutrino spectrum and adopted the best-fit values for the neutrino oscillation parameters as determined by the global fit \cite{deSalas:2020pgw} employing Casas-Ibarra parametrization. In the left panel of Fig.~\ref{fig:doubletDM}, the narrow band represents the $3\sigma$ experimental range of DM relic density $i.e.$ $\Omega h^{2}=(0.1126\leq\Omega h^2\leq0.1246$) taken from the Planck satellite experiment \cite{Planck:2018vyg}. The regions of under-abundant and over-abundant relic density are represented by cyan and yellow points, respectively. The orange-colored points are excluded by LEP constraints \cite{Belyaev:2016lok,Cao:2007rm}. The under-abundance is attributed to various dips, which result from different annihilation and co-annihilation channels. The first and second dips are due to the annihilation and coannihilation of $\eta_{R}$ mediated by the exchange of $W$ and $Z$ bosons, which results in a reduction of the DM relic density. The third dip occurs when the relic density decreases at half the mass of the Higgs boson ($h_{1}$), where $\eta_{R}$ annihilates $via$ Higgs exchange. At masses near 90 GeV, the quartic interactions between the DM particle and SM particles become significant. These interactions lead to an increase in the DM annihilation cross-section, particularly into W boson pairs. Consequently, the effective thermal freeze-out temperature rises, resulting in a reduction of the DM relic density. This effect is noticeable as a fourth dip in the relic density plot. Similarly, when the DM particle reaches a mass of approximately $125$ GeV, annihilation into Higgs boson pairs becomes a dominant channel due to the strengthened coupling between the DM particle and the Higgs boson. The blue points fall within the \( 3\sigma \) range of the cold DM relic density, as determined by the Planck collaboration measurements. These points are distributed across two distinct mass regions: the low mass range from 40 GeV to 85 GeV, and the high mass region from 550 GeV to approximately 4.5 TeV. We are not getting correct relic points after 4.5 TeV in the relic density plot. This is because the  loop corrections to the Higgs mass become
too large and cannot be counterbalanced by adjusting the tree-level Higgs-quartic ($\lambda_1$) coupling.\\
The right panel of Fig.~\ref{fig:doubletDM} illustrates the spin-independent direct detection cross-section of DM, where the interaction between DM and nucleons is mediated by the \( h_{1} \) and \( Z \)-bosons as shown in Fig.~\ref{fig:ddeta}. The experimental bound on CDM-nucleon cross-sections from XENONnT \cite{XENON:2023cxc}, PandaX-4T \cite{PandaX-4T:2021bab} and LZ \cite{LZ:2022lsv} are shown as distinct colored lines. The black line corresponds to the neutrino floor. 
The combination of all relevant constraints leads to an allowed mass range of $m_{\eta_R} \sim 40$ GeV $-$ 85 GeV and $m_{\eta_R} \sim 550$ GeV $-$ 4.5 TeV for scalar doublet DM. These results are consistent with those obtained in the canonical scotogenic model. However, they differ from the findings for scalar doublet dark matter in the Dirac Scotogenic model \cite{CentellesChulia:2024iom}, where co-annihilation between $\eta_R$ and $\eta_I$ is absent. In the upcoming sub-section, we will focus on fermionic DM in the model.
\subsection{Triplet Fermion Dark Matter ($\Sigma^0_{1}$)}
\noindent In the second scenario, we consider the $\mathcal{Z}_{2}$-odd neutral component of the
\begin{figure}[th]
    \centering
    \textbf{\Large $\Sigma^0_{1}$ as DM}\\
     \includegraphics[height=6.5cm]{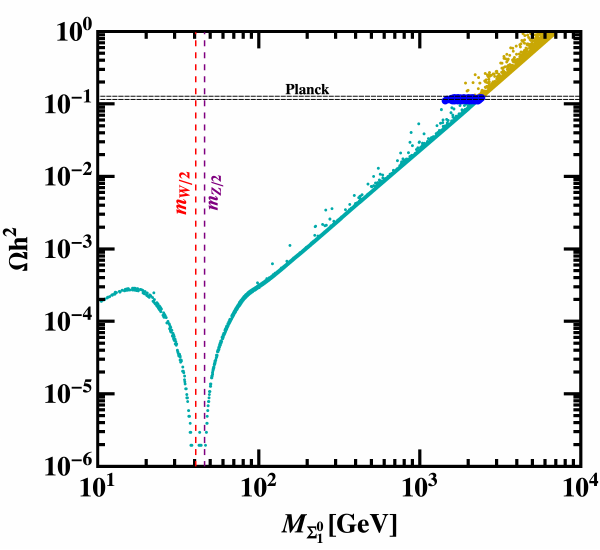}\\
     \includegraphics[height=6.5cm]{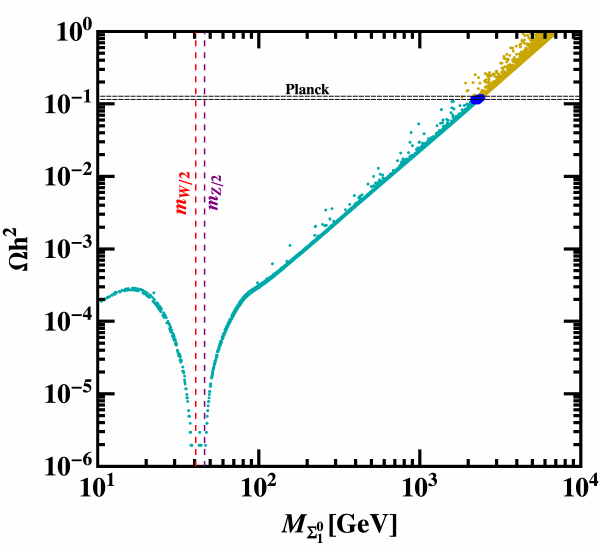}
     \includegraphics[height=6.5cm]{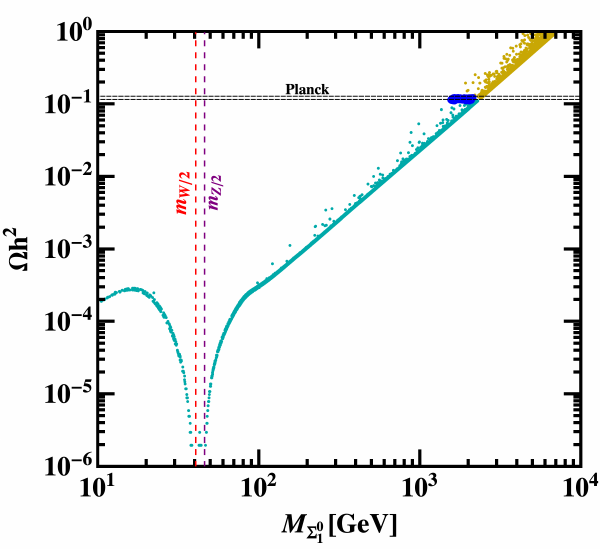}
    \caption{Relic density vs mass of the triplet fermion DM particle. \textbf{Top:} Considering both annihilation and co-annihilation channels, \textbf{Bottom Left:} Considering only annihilation channels, \textbf{Bottom Right:} Considering only co-annihilation channels. Color code is same as in Fig.~\ref{fig:doubletDM}.}
    \label{tripferm}
\end{figure} 
fermion triplet ($\Sigma^0_{1}$) as the DM candidate. To be definite, we choose $\Sigma^0_{1}$ as the lightest dark sector particle. This implies that $M_{\Sigma^0_1} < M_{\Sigma^0_2},M_{\Sigma^{\pm}_1},M_{\Sigma^{\pm}_2}$ and it is lighter than both the neutral and charged component of the scalar doublet ($M_{\Sigma^0_1} < m_{\eta_R}, m_{\eta_I}, m_{\eta^+}$).
Since the $\Sigma^0_1$ has a gauge invariant mass, we can easily take it to be lighter than other dark sector particles. Fig.~\ref{tripferm} illustrates the variation of the DM relic density with the mass of the triplet fermion. In the top panel, both annihilation and co-annihilation channels are considered, while the bottom left and bottom right panels focus on annihilation and co-annihilation channels separately. Regions of over-abundance and under-abundance are indicated by yellow and cyan colours, respectively. The annihilation and co-annihilation processes contributing to the DM relic density are detailed in Figs.~\ref{fig:ddeta1},~\ref{fig:ddeta2} and \ref{fig:ddeta3} in Appendix \ref{sec:appendix1}. A dip occurs at \(M_{DM} = M_{W, Z}/2\) due to annihilation processes mediated by the exchange of \(W\) or \(Z\) bosons.
The correct relic density is obtained at around $2.4$ TeV as shown in bottom left panel of Fig.~\ref{tripferm}, consistent with the findings reported in Ref.~\cite{Ma:2008cu}. It is important to note that Higgs mass constraints are not applicable here as triplet fermion do not couple with Higgs boson. Alternatively, we can consider the DM candidates to have nearly degenerate masses. In this case, co-annihilation processes with scalar doublet and charged components of triplet fermion ($\Sigma_{k}^\pm$)  play a crucial role in determining the relic density of the triplet neutral fermion. The inclusion of co-annihilation effects with other dark sector particles significantly broadens the available parameter space for triplet fermion DM, allowing masses to range from approximately $1.4$ TeV to $2.4$ TeV as shown in the top panel of Fig.~\ref{tripferm}. This mechanism provides a viable pathway to achieve the correct relic abundance within this extended mass range. Various co-annihilation channels contributing to lowering the  DM mass scale up to 1.4 TeV are summarized in Fig.~\ref{fig:ddeta3} in Appendix \ref{sec:appendix1}.

\begin{table}[t]
\begin{center}
\begin{tabular}{  |  c  |     c  |   c |  c |}
  \hline\hline
  Parameter    &   B1  &   B2 & B3  \\
\hline\hline
$\lambda_{1}$                 &  0.128 &  	 0.1275 & 0.1275   \\
$\lambda_{2}$       & $6.71 \times 10^{-3}$ &    $7.67 \times 10^{-3}$     &  $1.31 \times 10^{-3}$  \\
$\lambda_{3}$            & $5.65 \times 10^{-2}$ &    $9.5 \times 10^{-2}$ &  $0.06$   \\
$\lambda_{4}$          & $5.47 \times 10^{-2}$ &     $-7.4 \times 10^{-2}$  & $-3.37 \times 10^{-2}$   \\
$\lambda_{5}$             &  $-0.11$ &   $-1.5 \times 10^{-5}$ & $-1.54 \times 10^{-6}$  \\
$\mu^2_{\eta}[\text{GeV}^2]$      & $5.89 \times 10^{3}$ &      $3.36 \times 10^{5}$   &  $2.06 \times 10^{6}$  \\
$Y^\nu$       & $\begin{pmatrix}
    0.000007 & 0.000006 \\ 0.000006 & 0.000033 \\ -0.000008 & 0.000028
\end{pmatrix}$ &      $\begin{pmatrix}
    0.031 & 0.020 \\ 0.026 & 0.102 \\ -0.037 & 0.087
\end{pmatrix}$   &  $\begin{pmatrix}
    0.005 & 0.003 \\ 0.004 & 0.01 \\ -0.006 & 0.01
\end{pmatrix}$ \\
  $m_{h1}[\text{GeV}]$       & 125.29 &     125.14 	 &  125.24  \\	
	$m_{\eta_{R}}[\text{GeV}]$      & 76.24 &    578.99    &  1426.20 \\	
 	$m_{\eta_{I}}[\text{GeV}]$       & 112.89 &    578.99    &  1426.20 \\
	$m_{\eta^\pm} [\text{GeV}]$      & 87.48 &    581.27  &  1426.83  \\
  	$M_{\Sigma^0_{1}}[\text{GeV}]$      & 1497.90 &    1273.39    &  1424.70  \\
      	$M_{\Sigma^0_{2}}[\text{GeV}]$      & 4181.02 &    1650.59     &  2283.02  \\
         $M_{\Sigma_{1}^\pm}[\text{GeV}]$    & 1497.95 &    1273.44    &  1424.75  \\
        $M_{\Sigma_{2}^\pm}[\text{GeV}]$   & 4181.03  &    1650.63    &  2283.05 \\
 $\Omega h^2$           &  0.1143  &	  0.1239      &  0.1127  \\
 $\sigma^{SI} [cm^{2}]$               & $1.57 \times 10^{-48}$  &	 $6.28 \times 10^{-48}$ & 0 \\
    \hline
  \end{tabular}
\end{center} 
\caption{Relevant model  parameters for three representative benchmark points, B1 (low mass scalar doublet DM), B2 (medium mass scalar doublet DM) and B3 (fermion triplet DM).}
\label{tab:benchmark} 
\end{table}

\noindent It is important to note that as the mass difference between the two $\mathcal{Z}_{2}$-odd particles approaches zero, co-annihilation processes become more significant. As a result, the dark matter relic density is achieved at a slightly lower scale. In Table \ref{tab:benchmark}, we illustrate three benchmark points. For the sake of completeness, we have, also, calculated the branching ratio of LFV process $\mu\rightarrow e\gamma$  for both scalar doublet and triplet fermion DM case shown in Fig. \ref{fig:LFV} of Appendix \ref{appendixB}.

\section{Collider Phenomenology}\label{collider}
\noindent Numerous collider searches have been conducted to investigate the potential existence of the inert doublet \cite{Cao:2007rm,LopezHonorez:2006gr,Lundstrom:2008ai,Dolle:2009ft,Dolle:2009fn,LopezHonorez:2010eeh,Miao:2010rg,Gustafsson:2012aj,Arhrib:2012ia,Swiezewska:2012eh,Arhrib:2013ela,Krawczyk:2013jta,Belanger:2015kga,Poulose:2016lvz,Datta:2016nfz,Belyaev:2018ext}. As in any model with a DM candidate, the generic signature is missing energy $\cancel{E}$, measured from the total transverse momentum recoil of the visible particles in the events. The inert doublet of our model can be tested at LHC through a variety of signatures, including mono-jet, mono-Z, mono-Higgs and vector boson fusion + $\cancel{E}$ \cite{Arhrib:2013ela}. The inert scalar doublet of our model can also be explored at future high-energy electron-positron colliders, such as the Compact Linear Collider (CLIC) and the International Linear Collider (ILC). CLIC, a high-luminosity linear collider operating at the TeV scale, is designed to run at three center-of-mass energies: 380 GeV, 1.5 TeV, and 3 TeV \cite{Aicheler:2018arh}, with respective integrated luminosities of 1 ab\(^{-1}\), 2.5 ab\(^{-1}\), and 5 ab\(^{-1}\). Similarly, the ILC, planned to operate at a collision energy of 500 GeV \cite{Behnke:2013xla}, aims for an integrated luminosity of 4 ab\(^{-1}\), offering another robust platform to investigate the phenomenology of the inert scalar doublet of our model.

\begin{figure}[h!]
    \centering
      \includegraphics[height=4cm]{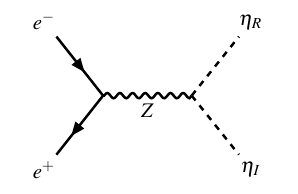}
            \includegraphics[height=4cm]{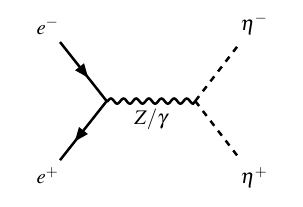}
    \caption{The Feynman diagrams for the process $e^- e^+ \rightarrow \eta_{R} \eta_{I}$ and $e^- e^+ \rightarrow \eta^+ \eta^-$.}
    \label{eeetr}
\end{figure}
\begin{figure}[t]
    \centering
     \includegraphics[height=6.5cm]{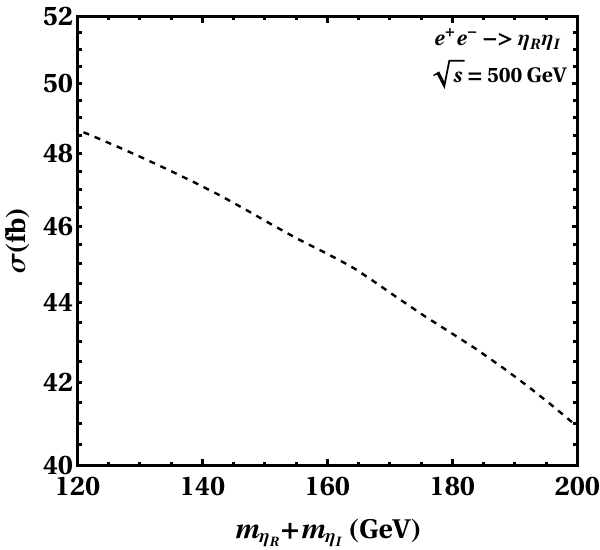}
    \includegraphics[height=6.5cm]{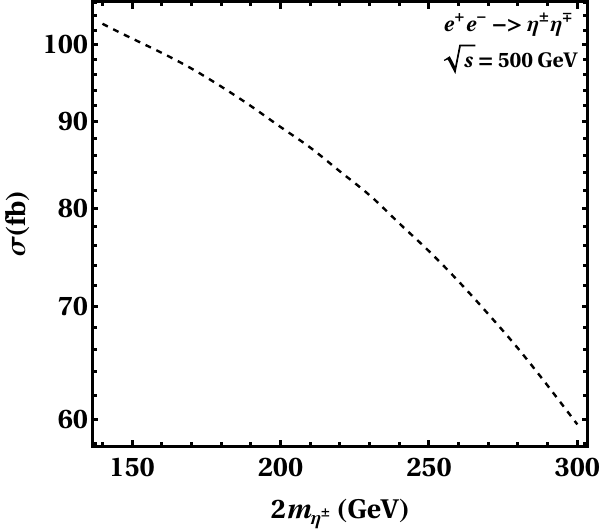}
    \caption{Production cross-section of pair production of the inert scalar doublet at ILC with centre-of-mass energy $\sqrt{s}=500$ GeV for BR1.}
    \label{ilc}
\end{figure} 

\noindent To calculate production cross-sections of the various processes, we first generated the model's UFO files using the SARAH  package \cite{Staub:2015kfa}, which then implemented into MadGraph5 3.5.4 version \cite{Alwall:2014hca} for further computation. To calculate the production cross-section, we examine two benchmark ranges (BR), BR1 and BR2, representing the low-mass and medium-mass regions of scalar DM, respectively, as shown in Fig.~\ref{fig:doubletDM} and detailed in Table \ref{tab:benchmark}. At lepton colliders, the pair production of $\eta_{R}\eta_{I}$ can be possible by $s$-channel ($t$-channel) exchange of $Z$-boson ($\Sigma^{\pm}_k$). The $s$-channel production of inert pairs is shown in Fig.~\ref{eeetr}. Also, the pair production of charged $\eta^{\pm}\eta^{\mp}$ can be possible by $s$-channel ($t$-channel) exchange of $Z$-boson ($\Sigma^{0}_k$). Fig.~\ref{ilc} presents the variation in cross-sections for the pair production of the processes $e^{+}e^{-} \rightarrow \eta_{R}\eta_{I}$ and $e^{+}e^{-} \rightarrow \eta^{\pm}\eta^{\mp}$ for centre of mass-energy $\sqrt{s}=500$ GeV for BR1\footnote{Note that certain points in these ranges are also ruled out by dark matter constraints as discussed in Sec.\ref{sec:DM}.} consistent with the LEP bound $i.e.$ $m_{\eta_R} + m_{\eta_I} > m_Z$ and $m_{\eta^+} > 70$. The $\eta^{\pm}$ decays predominantly into $W^{\pm}\eta_{R}$, where $W^{\pm}$ to be on-shell if $m_{\eta^{\pm}}-m_{\eta_{R}}>m_{W}$, which can further decay to quarks (jets) or/and mono or di-leptons. The $\eta_{R}$ predominantly decays to $Z\eta_{R}$, which can further decay di-leptonically.
\begin{figure}[h!]
 \centering
        \includegraphics[height=5.9cm]{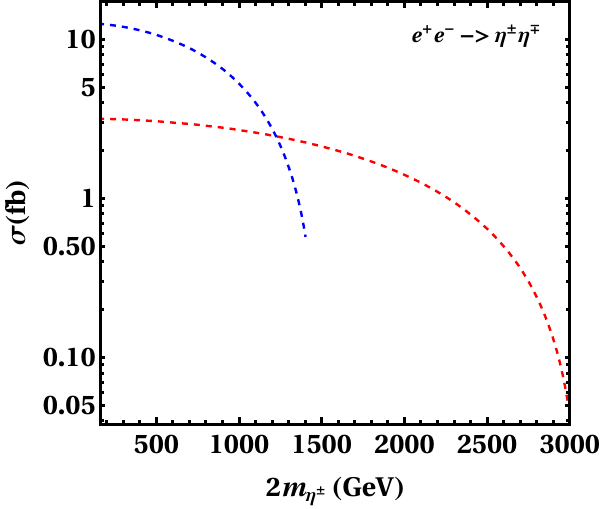}
        \includegraphics[height=5.9cm]{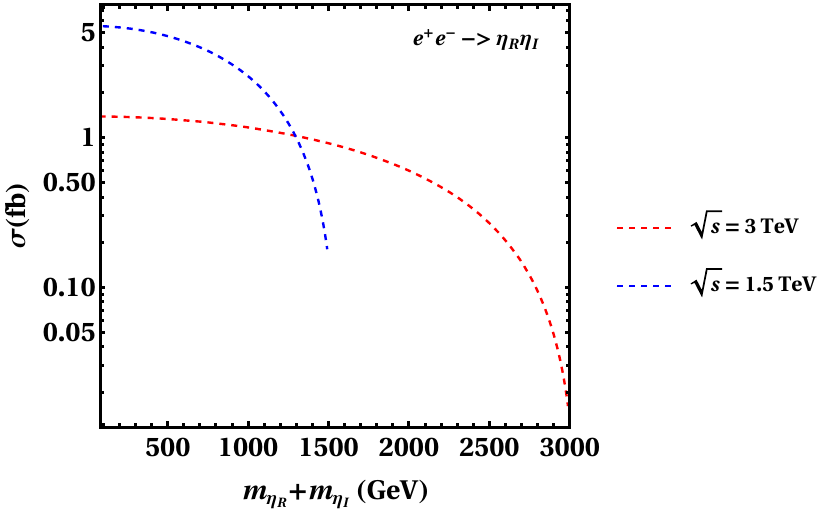}\\
         \includegraphics[height=5.9cm]{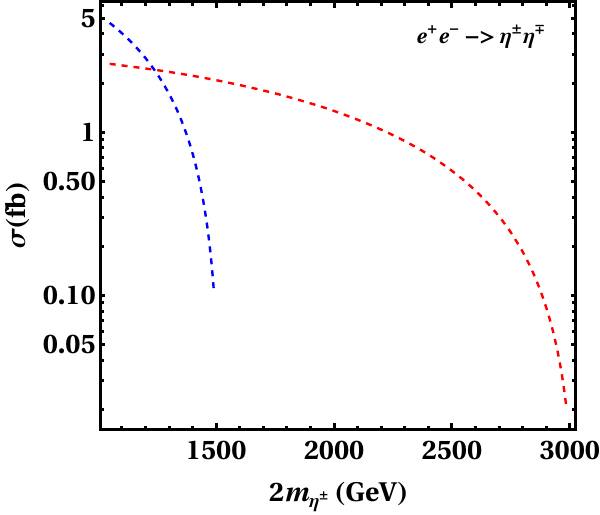}
        \includegraphics[height=5.9cm]{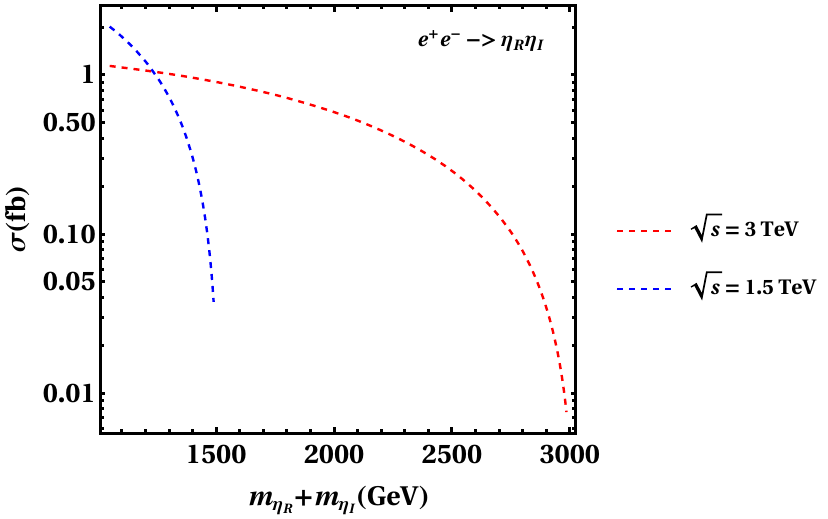}\\
        \caption{Production cross-section of the pair production of scalar doublet at lepton collider for BR1 (upper panel) and BR2 (lower panel).}
        \label{fig:lep1}
\end{figure}
Further, Fig.~\ref{fig:lep1} illustrates the production cross-sections for the processes $e^{+}e^{-} \rightarrow \eta^{\pm}\eta^{\mp}$ and $e^{+}e^{-} \rightarrow \eta_{R}\eta_{I}$ at centre-of-mass energies 1.5 TeV and 3 TeV of CLIC, shown for BR1 (upper panel) and BR2 (lower panel). The production cross-section is significantly larger for BR1 (low-mass DM region) compared to BR2 (medium-mass DM region) at the same center-of-mass energy. This difference arises because, at lower masses, the production process benefits from a larger available phase space, allowing for higher kinematic efficiency. Additionally, the reduced mass of the final-state particles in the BR1 scenario lowers the suppression from phase-space factors, further enhancing the cross-section. Conversely, in the BR2 case, the higher mass of the particles results in more constrained phase space, leading to a suppressed production rate. The various decay modes and the corresponding signatures of the particles are detailed in Table~\ref{sign}.
\begin{table}[t]
\begin{center}
\begin{tabular}{ | l | l  |r | }
  \hline \hline
  Process
& \hspace{0.1cm}  Decay mode  \hspace{0.1cm}          &  \hspace{0.1cm}  Signature               \\
\hline\hline
\multirow{4}{*}{\vspace{0.3cm} \begin{turn}{0} $e^+ e^- \rightarrow \eta^{\pm} \eta^{\mp} $ \end{turn} } &
 $\eta^{\pm} \rightarrow W^{\pm} \eta_R \rightarrow q_1 
 \Bar{q_2} \eta_R,  \eta^{\mp} \rightarrow W^{\mp} \eta_R \rightarrow q_3 
 \Bar{q_4} \eta_R$        	  &    4 jets + $\cancel{E_T}$    \\
 &   $\eta^{\pm} \rightarrow W^{\pm} \eta_R \rightarrow l^{\pm} 
 \nu_{l} \eta_R,  \eta^{\mp} \rightarrow W^{\mp} \eta_R \rightarrow q_3 
 \Bar{q_4} \eta_R$       &   1$l$ + 2 jets + $\cancel{E_T}$     \\
  &   $\eta^{\pm} \rightarrow W^{\pm} \eta_R \rightarrow l^{\pm}_1 
 \nu_{l} \eta_R,  \eta^{\mp} \rightarrow W^{\mp} \eta_R \rightarrow l^{\mp}_2 
 \nu_{l} \eta_R$       &   2$l$ + $\cancel{E_T}$     \\
  &   $\eta^{\pm} \rightarrow W^{\pm} \eta_R \rightarrow l^{\pm}_1 
 \nu_{l} \eta_R,  \eta^{\mp} \rightarrow l^{\mp}_2 \Sigma_1 \rightarrow l^{\mp}_2 \eta_R \nu_l$       &   2 $l$ + $\cancel{E_T}$     \\
&   $\eta^{\pm} \rightarrow W^{\pm} \eta_R \rightarrow q_1 
 \Bar{q_2} \eta_R,  \eta^{\mp} \rightarrow l^{\mp} \Sigma_1 \rightarrow l^{\mp} \eta_R \nu_l$       &   1$l$ + 2 jets +  $\cancel{E_T}$   \\

\hline 
\multirow{1}{*}{ \begin{turn}{0} $e^+ e^- \rightarrow \eta_{R} \eta_{I} $ \end{turn} } 
 &$\eta_{I} \rightarrow Z \eta_{R} \rightarrow  l^{\pm}_1 l^{\mp}_2 \eta_R $        	 &  2 $l$ + $\cancel{E_T}$     \\
\hline 
\multirow{1}{*}{ \begin{turn}{0}  $p p, e^+ e^- \rightarrow \Sigma^{\pm}_1 \Sigma^{\mp}_1 $ \end{turn} } &
  $\Sigma^{\pm}_1 \rightarrow \eta^{\pm} \nu_l \rightarrow \eta_R l^{\pm}_1 \nu_l, \Sigma^{\mp}_1 \rightarrow \eta^{\mp} \nu_l \rightarrow \eta_R l^{\mp}_2 \nu_l$  		 &  2 $l$ +  $\cancel{E_T}$      \\
  \hline 
\multirow{1}{*}{ \begin{turn}{0}  $p p \rightarrow \Sigma^{\pm}_1 \Sigma^0_1 $ \end{turn} } &
  $\Sigma^{\pm}_1 \rightarrow \eta^{\pm} \nu_l \rightarrow \eta_R l^{\pm} \nu_l, \Sigma^0_1 \rightarrow \eta_R \nu_l$  & 1 $l$ +  $\cancel{E_T}$    \\
    \hline
  \end{tabular}
\end{center}
\caption{Summary of various signatures of the inert doublet members and the lightest triplet fermion at the lepton colliders and FCC-hh.}
 \label{sign}
\end{table}
\begin{figure}[h!]
    \centering
     \includegraphics[height=6.5cm]{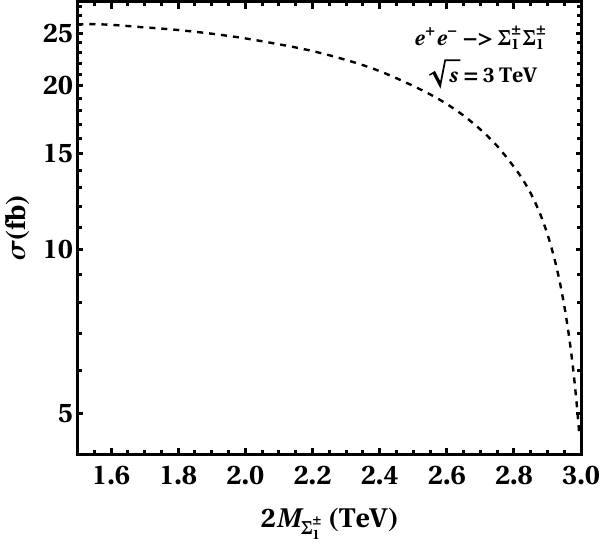}
    \caption{Production cross-section of pair production of charged triplet fermion at lepton collider CLIC with centre-of-mass energy $\sqrt{s}=3$ TeV for BR1.}
    \label{triplep}
\end{figure}
\begin{figure}[h!]
    \centering
     \includegraphics[height=6.2cm]{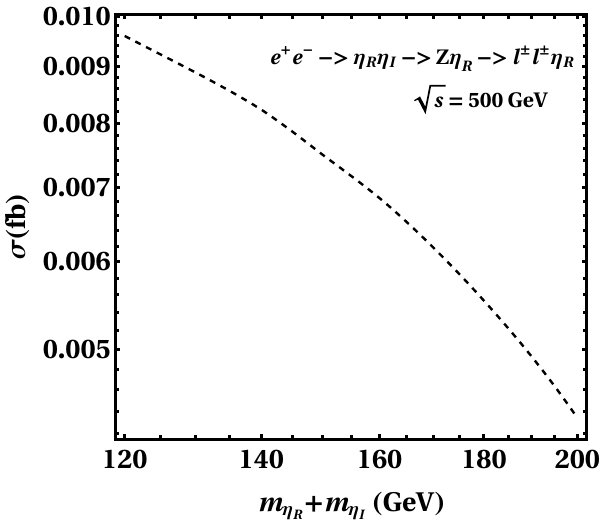}
     \includegraphics[height=6.2cm]{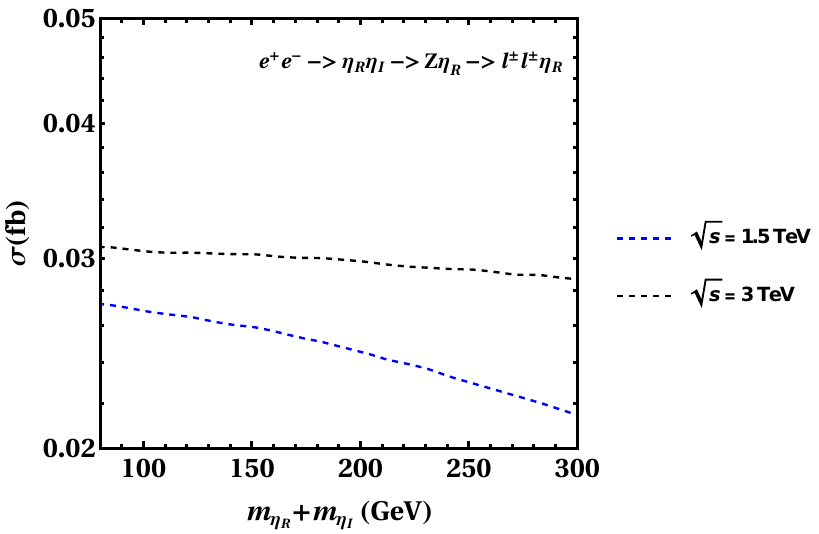}
     \includegraphics[height=6.2cm]{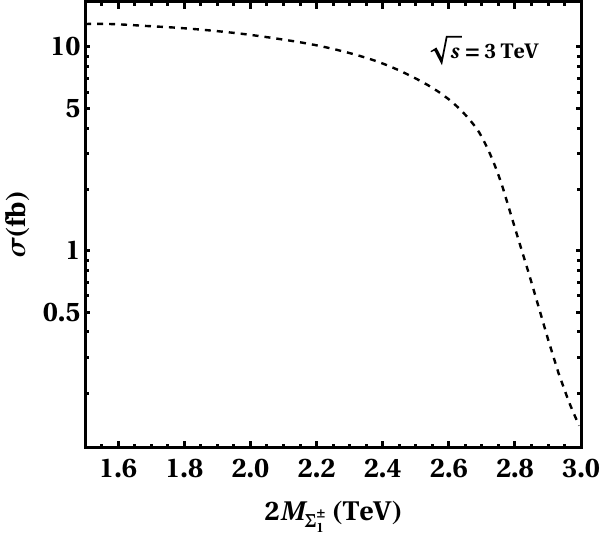}\\
    \caption{\textbf{Upper Panel:} Cross-section of di-leptonic decay process at ILC (\textbf{Left}) and CLIC (\textbf{Right}) for BR2. \textbf{Lower Panel:} The cross-section for the process $e^+ e^- \rightarrow \Sigma^{\pm}_1 \Sigma^{\mp}_1\rightarrow 2l+\cancel{E}_{T}$.}
    \label{chainrxn}
\end{figure}
Additionally, the production of charged components of the triplet fermion is possible at lepton colliders through $s$-channel ($t$-channel) $via$ exchange of $Z$-boson ($\eta_{I,R}$). The production of the triplet fermion is dominated by exchange of $Z$-boson because the $Y^{\nu}$ coupling is significantly small $i.e.$ $\mathcal{O}(10^{-5,-6})$. Fig.~\ref{triplep} shows the production cross-section of charged triplet fermion at centre-of-mass energy $\sqrt{s}=3$ TeV. The pair production of $\Sigma^{\pm}\Sigma^{\mp}$ would manifest as events with two oppositely charged lepton jets and neutrinos along with missing energy due to undetected neutral particles as shown in Table \ref{sign}. Further, the produced heavy particles can decay to the SM particles with missing energy and their collider signatures can be probed. The cross-section of a di-leptonic decay process with missing energy at ILC (CLIC), operating at a centre-of-mass energy of 500 GeV ( 1.5 TeV and 3 TeV) for BR1, is presented in Fig.~\ref{chainrxn} (upper panel). For this specific channel, the production cross-section is approximately $\mathcal{O}(10^{-2}\, \text{fb})$. The smallness of the cross-section is primarily attributed to phase space suppression, which significantly reduces the probability of the process occurring in the kinematic regime under consideration. The lower panel of Fig.~\ref{chainrxn} shows the cross-section for the process $e^+ e^- \rightarrow \Sigma^{\pm}_1 \Sigma^{\mp}_1\rightarrow 2l+\cancel{E}_{T}$ at 3 TeV. Here, the cross-section is $\mathcal{O}(10\, \text{fb})$.

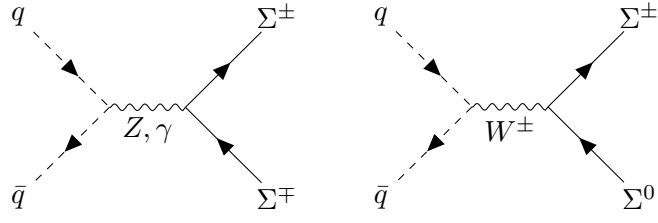
\begin{figure}[h!]
    \centering
   \begin{tikzpicture}
\begin{feynman}
\vertex at (0,0) (i1);
\vertex at (-1,0) (i2);
\vertex at (1,1) (a);
\vertex at (1,-1) (b);
\vertex at (-2,1) (c);
\vertex at (-2,-1) (d);

\vertex at (-2.2,-1.2) () {\(\bar{q}\)};
\vertex at (-2.2,1.2) () {\(q\)};
\vertex at (1.2,1.2) () {\(\Sigma^{\pm}\)};
\vertex at (1.2,-1.2) () {\(\Sigma^{\mp}\)};
\vertex at (-0.5,-0.3) () {\(Z,\gamma\)};
\diagram*{
(i2) -- [photon] (i1), (i1) -- [fermion] (a), (b) -- [fermion] (i1), (c) -- [charged scalar] (i2),(i2) -- [charged scalar] (d)
};
\end{feynman}
\end{tikzpicture}
\hspace{0.5cm}
\begin{tikzpicture}
\begin{feynman}
\vertex at (0,0) (i1);
\vertex at (-1,0) (i2);
\vertex at (1,1) (a);
\vertex at (1,-1) (b);
\vertex at (-2,1) (c);
\vertex at (-2,-1) (d);

\vertex at (-2.2,-1.2) () {\(\bar{q}\)};
\vertex at (-2.2,1.2) () {\(q\)};
\vertex at (1.2,1.2) () {\(\Sigma^{\pm}\)};
\vertex at (1.2,-1.2) () {\(\Sigma^{0}\)};
\vertex at (-0.5,-0.3) () {\(W^{\pm}\)};
\diagram*{
(i2) -- [photon] (i1), (i1) -- [fermion] (a), (b) -- [fermion] (i1), (c) -- [charged scalar] (i2),(i2) -- [charged scalar] (d)
};
\end{feynman}
\end{tikzpicture}
\caption{Feynman diagrams for the primary production mechanism for \(\Sigma^{+}\Sigma^{-}\) and \(\Sigma^{\pm}\Sigma^{0}\) pairs at the LHC also known as Drell-Yan process.}
\label{drellyan}
\end{figure}

\begin{figure}[h!]
 \centering
        \includegraphics[height=6.cm]{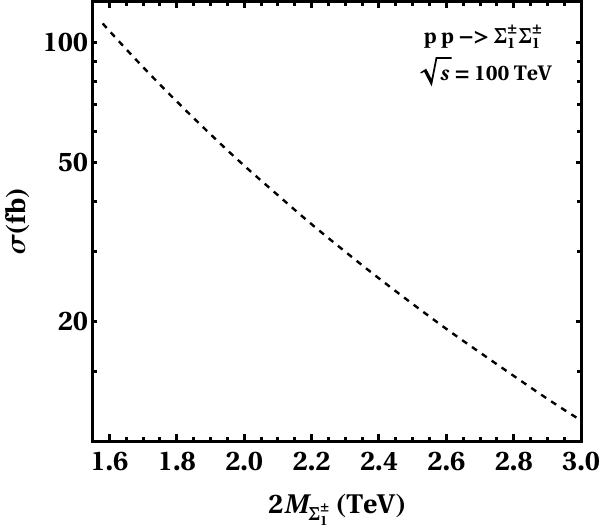}
        \includegraphics[height=6.cm]{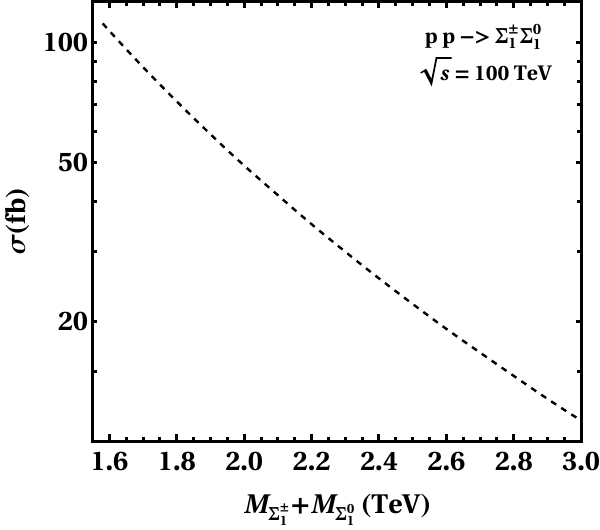}
        \caption{Production cross-section of triplet fermion through Drell-Yan process at FCC-hh with $\sqrt{s}=100$ TeV.}
        \label{pptrip}
\end{figure}

\begin{figure}[h!]
    \centering \includegraphics[height=6.2cm]{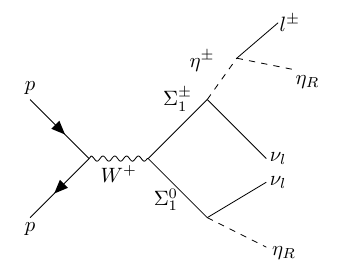}
     \includegraphics[height=6.5cm]{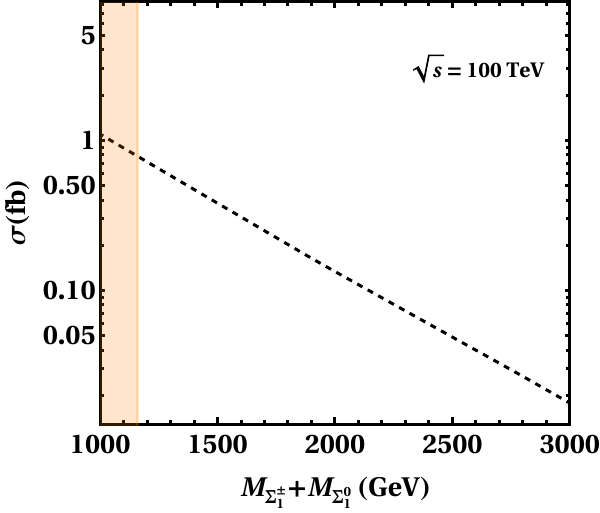}
    \caption{Feynman diagram (left) for the process 
$p p \rightarrow \Sigma_{1}^{\pm}\,\Sigma_{1}^{0} \rightarrow 1\ell + \cancel{E}_{T}$, and the corresponding cross section (right) at the future FCC-hh collider for the benchmark point BR1 in Table~\ref{tab:benchmark}. The shaded region denotes the parameter space excluded by current ATLAS bounds on triplet fermions given in Sec. \ref{lhc}.}
\label{smokinggun}
\end{figure}

\noindent The model can also be probed at the future circular collider for hadron-hadron collision (FCC-hh) \cite{FCC:2018vvp} with centre-of-mass energy $\sqrt{s}=100$ TeV. At FCC-hh, the pair production of charged component $\Sigma^{\pm}\Sigma^{\mp}$ is possible through $s$-channel $via$ exchange of $Z,\gamma$-boson, which is famously known as the Drell-Yan process. On the other hand, the pair production of $\Sigma^{\pm}\Sigma^{0}$ is possible through $s$-channel $via$ exchange of $W^{\pm}$ boson, as shown in Fig.~\ref{drellyan}.

\noindent If the triplet fermion is assumed to be a DM candidate, it satisfies the relic density requirements within the mass range of approximately 1.4–2.4 TeV. However, at such high masses, the production cross-section of the triplet fermion becomes exceedingly small, rendering it challenging to probe at the LHC with 13 TeV center-of-mass energy. Therefore, we utilize BR1 where scalar doublet is our DM candidate and we look for its collider signatures at future FCC-hh with center-of-mass energy 100 TeV. Fig.~\ref{pptrip} (upper panel)  shows the production cross-section of pair production of the process $p p \rightarrow\Sigma^{\pm}\Sigma^{\mp}$ (left panel) and $p p \rightarrow\Sigma^{\pm}\Sigma^{0}$ (right panel). While the production rates are relatively suppressed for higher masses due to phase-space limitations and the Yukawa coupling's dependence on mass, these processes remain crucial for investigating the triplet fermion sector. Further, the triplet fermion can decay $\Sigma^{\pm}_1 \rightarrow \eta^{\pm} \nu_l \rightarrow \eta_R l^{\pm} \nu_l, \Sigma^0_1 \rightarrow \eta_R \nu_l$ giving mono-leptonic and di-leptonic signatures at LHC as shown in Table \ref{sign}. The pair produced $\Sigma^{\pm}\Sigma^{0}$ can further decay mono-leptonically with missing energy. Fig.~\ref{pptrip} (lower panel) depicts the cross-section of mono-leptonic decay of pair-produced $\Sigma^{\pm}\Sigma^{0}$. It is worth mentioning that the $1\ell + \cancel{E}_T$ signature (given in Table \ref{sign}) arises dominantly from processes such as $pp \to \Sigma_1^\pm \, \Sigma_1^0 \to \ell^\pm \, \eta_R \, \nu_\ell$ (Feynman diagram given in Fig. \ref{smokinggun} (left panel)). The presence of a single isolated charged lepton accompanied by substantial missing transverse energy is a distinctive prediction when fermion triplets decay via inert scalars. Unlike multi-lepton or purely hadronic final states common in other BSM scenarios, the $1\ell + \cancel{E}_T$ channel here directly links the production of heavy triplet fermions to the dark sector, making it a smoking gun for this radiative neutrino mass framework. Fig. \ref{smokinggun} (right panel) depicts the cross-section of the $1\ell + \cancel{E}_T$ process at FCC-hh collider for the BR1 given in the Table \ref{tab:benchmark}. Future high-energy collider FCC-hh, with enhanced luminosities and collision energies, could significantly improve the sensitivity to such signatures, offering a robust test of the triplet fermion's role in DM phenomenology.

\section{Conclusions}\label{conclusion}
\noindent We have extended the Type-III scotogenic neutrino mass model by incorporating a real singlet scalar field. This enhancement enables the model to address challenges associated with Higgs inflation while simultaneously providing a viable framework for $cold$ cosmic inflation. The inflationary mechanism operates analogously to Higgs inflation. In our framework, the real singlet scalar remains effectively decoupled from the electroweak sector owing to its large mass and extremely suppressed couplings 
$(\lambda_7, \lambda_8 \sim 10^{-6})$, thereby leaving neutrino mass generation, dark matter stability, and electroweak vacuum stability unchanged. The inflaton mass is set to $\mathcal{O}(10^{8}~\mathrm{GeV})$, with a reheating temperature of $\mathcal{O}(10^{11}~\mathrm{GeV})$, ensuring efficient post-inflationary production of heavy inert scalars and fermion triplets in the dark sector. This choice guarantees that inflaton decays into TeV - scale dark-sector particles are kinematically allowed, while the high reheating temperature sustains their unsuppressed production, consistent with the observed relic abundance. 
Consequently, although the singlet does not influence low-energy collider or neutrino observables, it serves as the crucial link between early-Universe inflationary dynamics and the viable dark matter production mechanism in the model. The dark sector of the model consists of two DM candidates: the real part of the neutral component of the scalar doublet and the neutral component of the triplet fermion. For the scalar doublet DM case, two distinct mass regions are viable: a low mass region near half Higgs mass and a medium mass region ranging from 550 GeV - 4.5 TeV. In the triplet fermion case, the correct relic density is obtained at around 2.4 TeV by considering annihilation channels. However, if the mass of other dark sector particles is close to the mass of the triplet neutral fermion, in which case the co-annihilation channels become important, opening up an extra parameter space ranging from 1.4 TeV - 2.4 TeV.\\ 
\noindent We, also, discuss the collider phenomenology of the model by presenting the cross-sections for various processes corresponding to two benchmark ranges: BR1, representing low-mass scalar dark matter, and BR2, representing medium-mass scalar dark matter, which satisfies all the theoretical constraints on the model. We analyze the model for both leptonic and hadronic colliders. Various decay modes and respective signatures at colliders are shown in Table \ref{sign}. The $e^{+}e^{-}$ colliders, such as ILC and CLIC, are ideal for precise studies of weakly interacting particles and processes like $e^{+}e^{-} \rightarrow \eta^{+}\eta^{-}$ and $e^{+}e^{-} \rightarrow \eta_{R}\eta_{I}$ can be studied with minimal background noise. Given the small production cross-section of the triplet fermion at the LHC, we evaluated our model at FCC-hh and find that the mono-leptonic decay of the triplet fermion, accompanied by missing energy, is a significant signature, as shown in Fig.~\ref{smokinggun}. We find that the leading order production cross-section of low-mass scalar dark matter can be probed at both leptonic and hadronic colliders. \\

\section{Acknowledgments}
\vspace{.3cm}
\noindent LS acknowledges the financial support provided by the Council of Scientific and Industrial Research (CSIR) vide letter No. 09/1196(18553)/2024-EMR-I. SY acknowledges the funding support by the CSIR NET-SRF fellowship.

\appendix

\section{Feynman diagrams for annihilation/co-annihilation, production and detection of DM}
\label{sec:appendix1}

\begin{figure}[h!]
        \includegraphics[width=14cm]{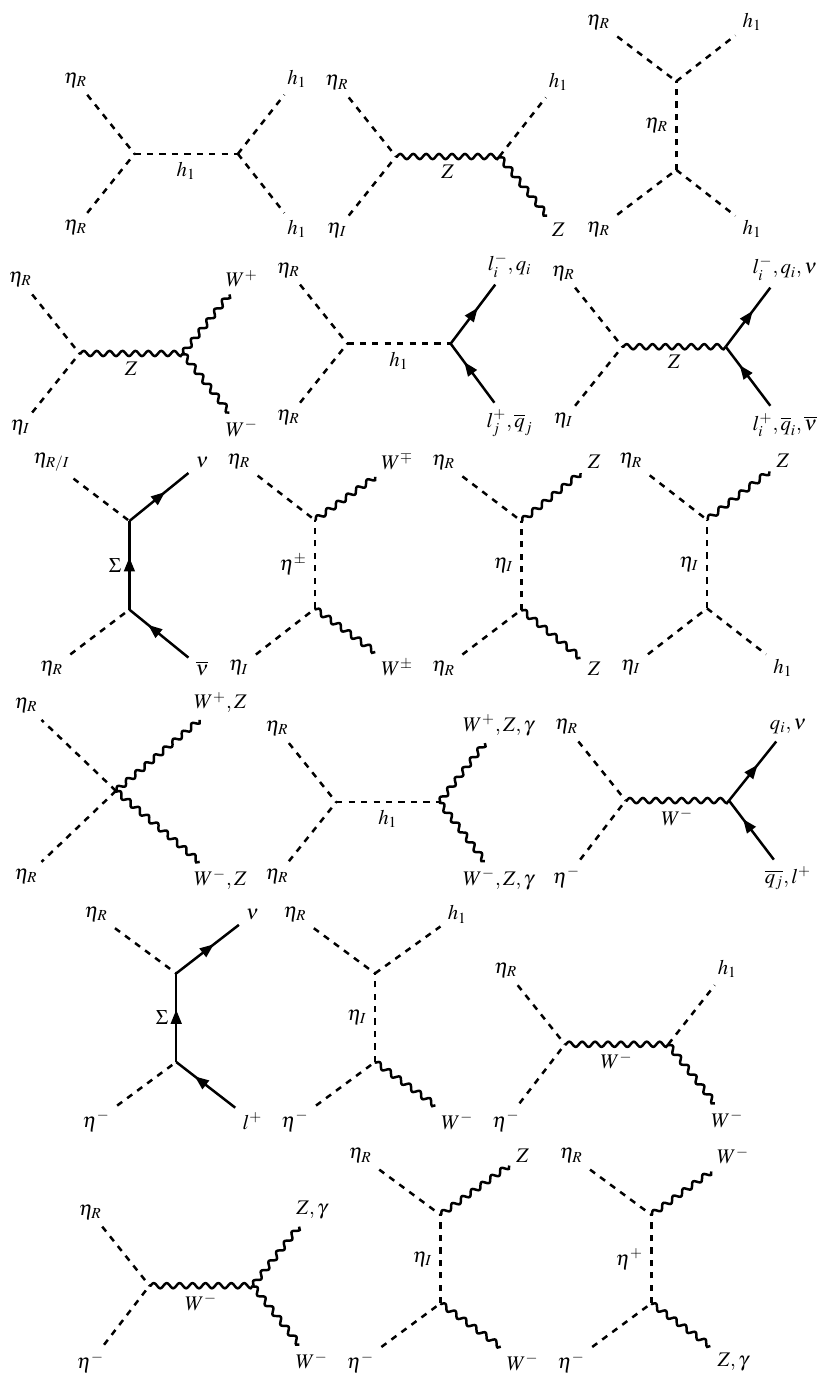}
        \caption{Relevant diagrams (annihilation/co-annihilation channels) for computing the relic density of doublet scalar DM candidate.}
                \label{fig:relicEta}
\end{figure}

\begin{figure}[h!]
       \includegraphics[height=5cm]{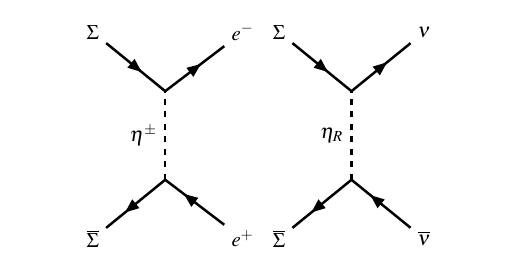}
        \caption{Annihilation channels for computing the relic density of triplet fermion DM candidate.}
        \label{fig:ddeta1}
\end{figure}

\begin{figure}[h!]
        \includegraphics[height=13cm]{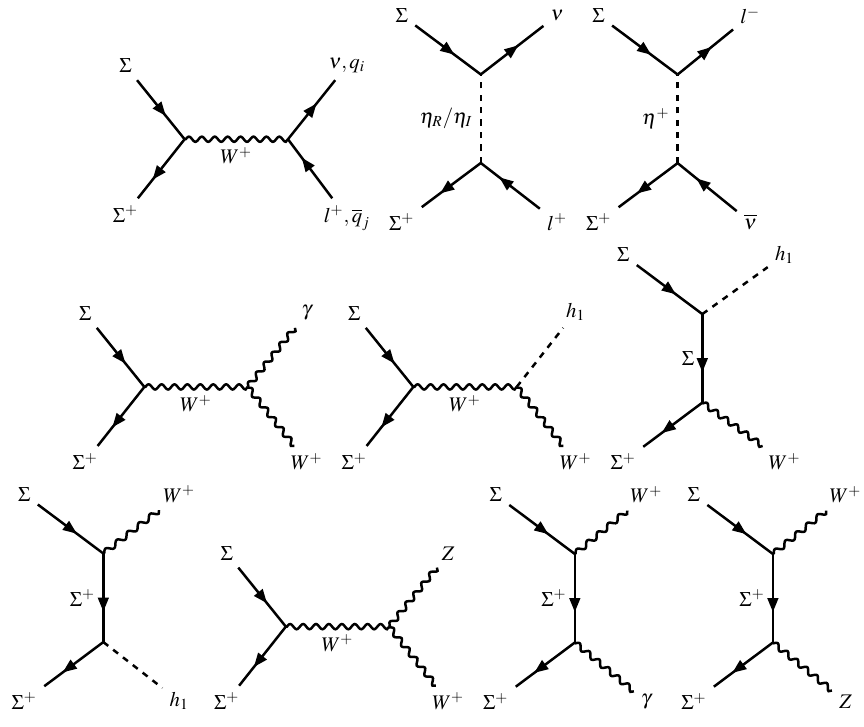}
        \caption{Co-annihilation channels (with $\Sigma^{+}$) for computing the relic density of triplet fermion DM candidate.}
                \label{fig:ddeta2} 
\end{figure}

\begin{figure}[h!]
        \includegraphics[height=17cm]{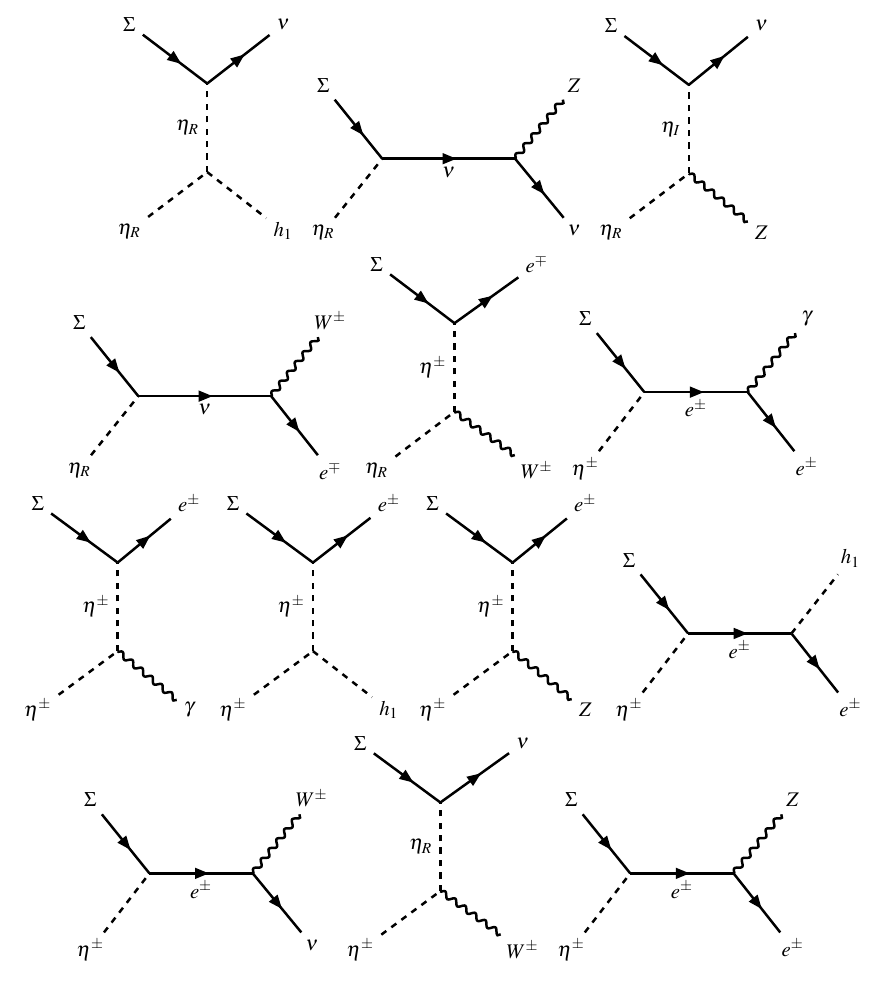}
        \caption{Co-annihilation channels for computing the relic density of triplet fermion DM candidate.}
                \label{fig:ddeta3} 
\end{figure}

\begin{figure}[h!]
        \includegraphics[height=5cm]{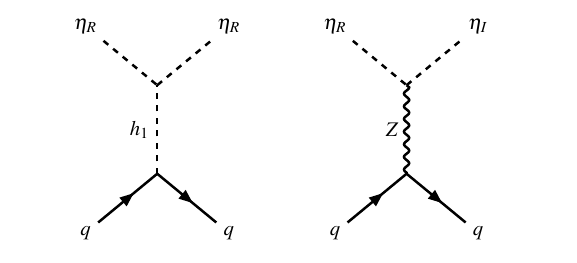}
        \caption{Relevant diagrams for the direct detection of the doublet scalar DM candidate.}
                \label{fig:ddeta} 
\end{figure}

\noindent In Figs.~\ref{fig:relicEta} to \ref{fig:ddeta3}, we list the possible diagrams for production/annihilation of DM, relevant in the early universe, for the cases in which the DM is a doublet scalar or a triplet fermion, respectively. In  Fig.~\ref{fig:ddeta}, we show the direct detection prospects of the scalar doublet DM by exchange of a Higgs or $Z$ bosons.

\FloatBarrier

\section{LFV Plots}\label{appendixB}
\begin{figure}[h!]
        \includegraphics[height=7cm]{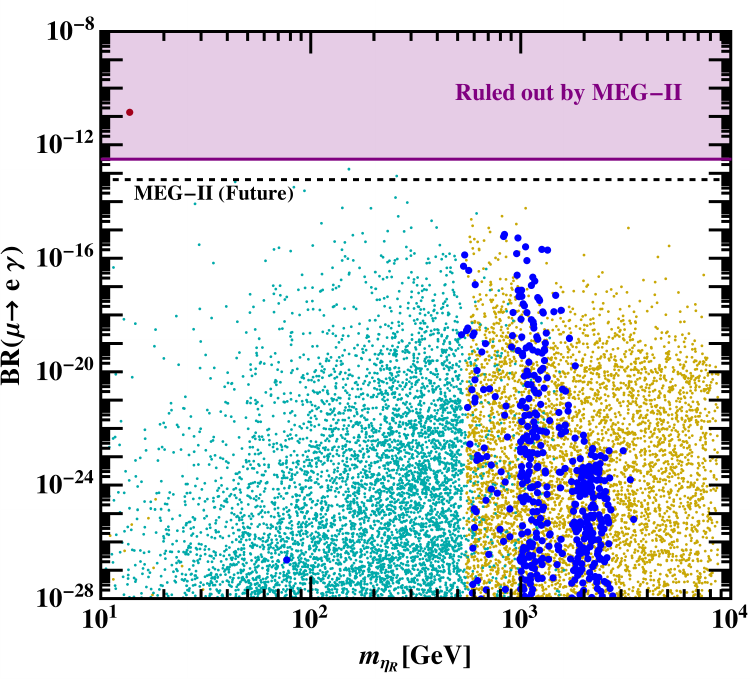}
         \includegraphics[height=7cm]{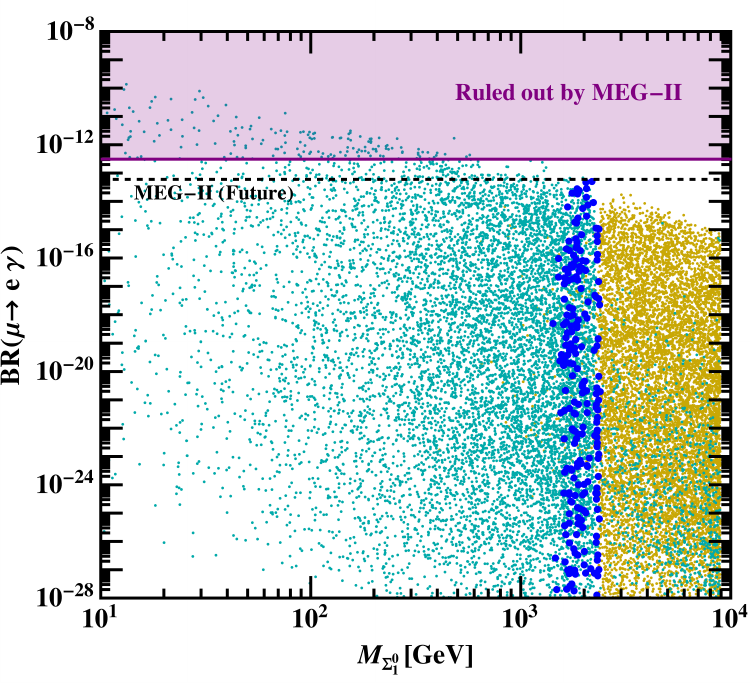}
        \caption{BR($\mu \rightarrow e \gamma$) vs doublet scalar DM mass (Left panel) and triplet fermion mass (Right panel).}
                \label{fig:LFV} 
\end{figure}

\noindent In Fig.~\ref{fig:LFV}, we show the branching ratio for $\mu \rightarrow e \gamma$ decay for both scalar doublet and triplet fermion DM case. For comparison, current limits at 90$\%$ C.L. are also shown in the figure as a solid purple line \cite{MEGII:2023ltw}, as well as the expected future sensitivity as dashed black line \cite{MEGII:2021fah}. From Fig.~\ref{fig:LFV}, it is clear that these rates are below the current experimental bound for correct relic density points.

\bibliography{bibliography.bib}
\end{document}